\documentclass[
reprint,
aps,
prb,
]{revtex4-2}
\usepackage{caption}
\usepackage{mathrsfs}
\usepackage[utf8]{inputenc}
\usepackage{graphicx}
\usepackage{amsmath,amssymb}
\usepackage{booktabs}
\usepackage{float}
\usepackage{etoolbox}
\usepackage{xcolor}
\usepackage{multirow}
\usepackage{amsmath}
\usepackage{titlesec}
\titlespacing{\section}{0pt}{*1.2}{*0.8}

\usepackage[font=small,skip=4pt]{caption}  
\makeatletter
\patchcmd{\@startsection}
  {\@afterheading}
  {\vspace*{-0.8em}\@afterheading}
  {}{}
\makeatother
\begin{document}

\title{Exchange-Driven Chiral Magnons and Weyl States in\\ Room-Temperature Metallic XCoB$_2$ (X= Ta, Zr and Hf) Altermagnets}

\author{Arafat Rahman}
\author{Tareq Mahmud}
\email{tareqphy1205@gmail.com}
\author{Alamgir Kabir}
\email{alamgir.kabir@du.ac.bd}
\affiliation{Department of Physics, University of Dhaka, Dhaka 1000, Bangladesh}

\begin{abstract}Metallic altermagnets remain rare, particularly in low-symmetry three-dimensional crystals where spin-split electronic bands and chiral magnon excitations can coexist. This work reports a family of metallic $d_{yz}$-wave altermagnets in the orthorhombic ternary borides XCoB$_2$ (X = Ta, Zr, Hf), which crystallize in the centrosymmetric $Pnma$ structure with G-type collinear magnetic order. Symmetry analysis within the magnetic space group $Pnm'a'$ (BNS No.~62.447) predicts a nonrelativistic spin splitting proportional to $k_yk_z$, with symmetry-enforced degeneracy on the $k_y$ and $k_z$ nodal planes, in excellent agreement with first-principles calculations. On the $k_x=0$ plane, the momentum-averaged spin splitting reaches 88.4, 52.3, and 70.6~meV at the Fermi level in TaCoB$_2$, ZrCoB$_2$, and HfCoB$_2$, respectively, with maxima exceeding 200~meV in all three compounds. Exchange analysis shows that the altermagnetic magnon splitting originates from symmetry-inequivalent sixth-neighbour inter-sublattice interactions. The chirality splitting reaches 1.51, 2.03, and 3.10 meV below 50 meV in TaCoB$_2$, ZrCoB$_2$, and HfCoB$_2$, respectively, making it accessible to inelastic neutron scattering. Monte Carlo simulations yield N'eel temperatures of $67\pm1.1$, $330\pm5.7$, and $307\pm5.0$~K, placing ZrCoB$_2$ and HfCoB$_2$ above room temperature. With spin-orbit coupling included, all three compounds host symmetry-protected Weyl points near the Fermi level, associated Fermi-arc surface states, and sizable intrinsic anomalous Hall conductivities $\sigma_{zx}$ of $+342$, $-392$, and $-221$~S/cm at the Fermi level, reaching maximum magnitudes of $763$, $942$, and $924$~S/cm for TaCoB$_2$, ZrCoB$_2$, and HfCoB$_2$, respectively. $X$CoB$_2$ therefore provides a single compensated platform carrying both magnonic and electronic chirality, one in the spin waves and the other in the Berry curvature, without any stray field.
\end{abstract}

\maketitle

\section{Introduction}
Exchange interactions govern the relative alignment of spins and the resulting magnetic order in solids~\cite{Anderson1950,Anderson1959,Liechtenstein1987}. Collinear magnetism is traditionally classified as ferromagnetism, with finite net magnetization, spin-split bands, and stray fields, or antiferromagnetism, where symmetry-related compensated sublattices enforce spin degeneracy, $E_\uparrow(\mathbf{k})=E_\downarrow(\mathbf{k})$, throughout the Brillouin zone~\cite{Jungwirth2016AFM,Baltz2018AFM}. More recently, altermagnetism has emerged as a third class of collinear magnetic order that combines the zero net magnetization of antiferromagnets with the spin-split electronic bands commonly associated with ferromagnets~\cite{Smejkal2022a}. The key distinction comes from symmetry. In an altermagnet, the opposite-spin sublattices are related neither by a fractional translation nor by inversion, but by a proper or improper crystallographic rotation~\cite{Smejkal2022b}. This symmetry preserves magnetic compensation while lifting spin degeneracy at general momenta, with degeneracy remaining on symmetry-protected nodal planes. The resulting nonrelativistic spin splitting (NRSS) survives without spin-orbit coupling (SOC), can reach the eV scale, and shows characteristic $d$-, $g$-, or $i$-wave momentum dependence~\cite{Smejkal2022a,Smejkal2022b,Hayami2019,Ahn2019,Yuan2020Fluorides,Mazin2021}. These magnetic states can be classified using spin groups, in which spin and lattice operations are treated independently, a framework introduced decades ago~\cite{Litvin1974} and recently developed into a systematic classification of magnetic crystals~\cite{Liu2022SpinGroups,Smejkal2022a,belov1957shubnikov,gallego2016magndata}.
\\
The NRSS of electronic bands is now well established experimentally. Angle-resolved photoemission has revealed spin splittings of several hundred meV in hexagonal $\alpha$-MnTe ($P6_3/mmc$, $g$-wave, $T_\mathrm{N}=307$~K)~\cite{Krempasky2024,Lee2024,Osumi2024}, a plaid-like splitting in pyrite MnTe$_2$~\cite{Zhu2024MnTe2}, and a large splitting in metallic NiAs-type CrSb, which orders near 700~K~\cite{Reimers2024CrSb}. Other proposed candidates include doped FeSb$_2$~\cite{Mazin2021}, layered V$_2$Se$_2$O~\cite{Ma2021}, organic charge-transfer salts~\cite{Naka2019}, and low-$Z$ fluoride antiferromagnets~\cite{Yuan2020Fluorides}, while rutile RuO$_2$, once considered a prototype metallic altermagnet~\cite{Ahn2019,Feng2022RuO2}, has recently been reported to have a nonmagnetic bulk ground state~\cite{Kessler2024RuO2,Hiraishi2024}. Broken time-reversal symmetry allows anomalous Hall responses without net magnetization~\cite{vsmejkal2020crystal,Smejkal2022b,Song2025NRM}, spin-split bands can support giant and tunneling magnetoresistance without stray fields~\cite{Smejkal2022TMR}, and crystal-axis-dependent spin currents are suitable for field-free switching~\cite{GonzalezHernandez2021,Bai2023PRL}. However, useful metallic altermagnets remain scarce. In high-symmetry systems such as CrSb and MnTe, symmetry can equalize the group velocities of the two spin channels, leaving the longitudinal current unpolarized despite strong band splitting~\cite{Jiang2025KV2Se2O}. Lower-symmetry three-dimensional metallic altermagnets that can support longitudinal spin-polarized currents, particularly those ordering above room temperature, are therefore of considerable interest~\cite{Reimers2024CrSb,Yang2025CrSb,Zeng2024CrSb,Jiang2025KV2Se2O,Zhang2025RbV2Te2O,Reichlova2024Mn5Si3}.
\\
Altermagnetism also has a direct magnonic counterpart, since the alternating exchange responsible for electronic spin splitting can lift the degeneracy of magnon branches with opposite chirality~\cite{Smejkal2023chiral,McClarty2024,Corticelli2022}. Their characteristic energy scale is governed mainly by exchange rather than magnetic anisotropy or an applied field, allowing excitations in the THz regime, well above the GHz frequencies typical of conventional ferromagnetic magnonics~\cite{Chumak2015,Barman2021,Nemec2018,Han2023}. Chiral magnon splitting has been observed by inelastic neutron scattering in $\alpha$-MnTe, where a 2~meV splitting is associated with inequivalent $J_{10}$ and $J_{11}$ superexchange pathways~\cite{Liu2024MnTe}, and in $\alpha$-Fe$_2$O$_3$, where a 3~meV splitting near 100~meV is linked to thirteenth-neighbor couplings~\cite{Sun2025Hematite,Hoyer2025Hematite}. The rutile difluorides show a broader range of behavior, from an undetectably small splitting in MnF$_2$~\cite{MnF2_absence_2025} and competition with dipolar effects in FeF$_2$~\cite{Sears2026FeF2} to a predicted meV-scale splitting in CuF$_2$ driven by resonance-enhanced Cu-F$\cdots$F-Cu super-superexchange and inequivalent $J_{7a}$ and $J_{7b}$ interactions~\cite{Ho2026CuF2}. Chiral magnon spectra have also been predicted for CrSb~\cite{Zhang2025CrSb} and strain-tuned $\beta$-MnO$_2$~\cite{JhaMnO2_2026}, with circularly polarized resonant inelastic x-ray scattering as a complementary probe~\cite{Biniskos2025RIXS}. These results show that magnon splitting is controlled by weak, long-range, symmetry-inequivalent exchange pathways rather than the dominant nearest-neighbor interactions~\cite{Liu2024MnTe,Ho2026CuF2,Goodenough1955,Kanamori1959}. Because the split branches carry opposite angular momentum, their spin-current polarization can be selected by frequency or momentum without an external magnetic field, providing opportunities for magnon spin Seebeck, spin Nernst, and orbital Nernst effects~\cite{Cui2023,Weissenhofer2024,MagnonOrbitalNernst2026}, as well as routes toward picosecond electrical switching, coherent THz excitation, sub-THz spin-current generation, and broadband spintronic THz emission~\cite{Olejnik2018,Kampfrath2011,Li2020,Vaidya2020,Seifert2016}. These electronic and magnonic developments and their device prospects have been reviewed recently~\cite{Bai2024AFM,Song2025NRM,Jungwirth2025NP}. Metallic altermagnets in the orthorhombic $Pnma$ structure remain comparatively unexplored, despite the identification of altermagnetic order in several $Pnma$ compounds~\cite{Rooj2025,NaOsO3_2025,Galindez2025}.
\\
A further direction is the coexistence of altermagnetism with topological semimetallic states. In such systems, the nodes can carry spin, producing spin-polarized surface Fermi arcs and strongly anisotropic spin transport that are absent in spin-degenerate antiferromagnetic Weyl semimetals~\cite{Yang2025AMWeyl,HelicalArc2025}. They can also exhibit a ferromagnet-like Hall response without net magnetization, because the intrinsic anomalous Hall conductivity is governed by the integrated Berry curvature rather than by the magnetic moment~\cite{Reichlova2024Mn5Si3,Liu2018Co3Sn2S2,StrainBerryMnTe2025,Zhou2024CrystalThermal,armitage2018weyl,wan2011topological,xiao2010berry,nagaosa2010}, together with chiral-anomaly magnetotransport~\cite{Huang2015ChiralAnomaly} and a large anomalous Nernst response~\cite{Ikhlas2017ANE}. Combined with chirality-split THz magnons, a single compensated material could therefore host both magnonic and electronic chirality, carried respectively by spin waves and Berry curvature, without stray fields. The known compensated Weyl systems are noncollinear or ferrimagnetic rather than altermagnetic~\cite{Nakatsuji2015Mn3Sn,Shi2018Ti2MnAl}, while altermagnetic Weyl states have been confirmed experimentally only in the hexagonal CrSb family~\cite{Li2025WeylCrSb,Lu2025FermiArcsCrSb} and predicted in intercalated transition-metal dichalcogenides~\cite{Sah2026TMD}. Collinear altermagnets in lower-symmetry three-dimensional crystal classes that simultaneously host symmetry-protected Weyl nodes, a resolvable chiral magnon splitting, and above-room-temperature magnetic order have not, to our knowledge, been reported.
\\
In this work, we show that the orthorhombic ternary borides $X$CoB$_2$ ($X=$ Ta, Zr, Hf), crystallizing in the centrosymmetric $Pnma$ structure, form a family of metallic altermagnets. Using symmetry analysis and first-principles calculations, we establish their altermagnetic character and investigate the resulting NRSS, exchange interactions, and chiral magnon spectra. We identify the microscopic exchange mechanism responsible for the magnon splitting and determine the magnetic ordering temperatures using Monte Carlo simulations. Our results establish $X$CoB$_2$ as a platform for studying coupled electronic and magnonic responses in metallic altermagnets.
\section{Methodology}
First-principles calculations were performed within density functional theory (DFT) using the projector augmented-wave (PAW) method as implemented in \textsc{vasp}~\cite{kresse1996a,kresse1996b,blochl1994,kresse1999}. The exchange-correlation interaction was described within the generalized gradient approximation (GGA) using the Perdew--Burke--Ernzerhof (PBE) functional~\cite{pbe1996}. The PAW potentials employed valence configurations of $5p^{6}5d^{3}6s^{2}$ for Ta, $4s^{2}4p^{6}4d^{2}5s^{2}$ for Zr, $5p^{6}5d^{2}6s^{2}$ for Hf, $3d^{8}4s^{1}$ for Co, and $2s^{2}2p^{1}$ for B. A plane-wave kinetic-energy cutoff of 600~eV was used throughout. The Brillouin zone was sampled using $\Gamma$-centered Monkhorst--Pack meshes~\cite{monkhorst1976} of $8\times4\times3$ for self-consistent calculations and $16\times8\times6$ for the non-self-consistent calculations used for Wannierization. Gaussian smearing of 0.05~eV was employed, with an electronic convergence criterion of $10^{-6}$~eV, while structural relaxation was continued until the residual forces were below $0.02$~eV\,\AA$^{-1}$. The Co $3d$ states were treated within the rotationally invariant DFT+$U$
	approach of Dudarev \textit{et al.}~\cite{dudarev1998}. We used an effective
	Coulomb parameter $U_{\mathrm{eff}}=U-J=3.0$~eV, a value within the range
	commonly adopted for cobalt in DFT+$U$ studies~\cite{SantosCarballal2021,Long2020}.\\
The magnetic ground state was initialized in the $G$-type antiferromagnetic configuration with alternating Co moments. Spin--orbit coupling (SOC) was included through fully non-collinear calculations~\cite{steiner2016}. Phonon dispersions were calculated using density functional perturbation theory (DFPT)~\cite{baroni2001,gajdos2006} with a $3\times2\times2$ supercell and processed using \textsc{phonopy}~\cite{togo2015}. Maximally localized Wannier functions (MLWFs) were constructed from $X(d)$ ($X=\mathrm{Ta}$, Zr, and Hf), Co$(d)$, and B$(p)$ orbital projections using \textsc{wannier90}~\cite{Pizzi2020}, interfaced with \textsc{vasp}, following the standard localization procedure~\cite{marzari1997,souza2001}.\\Berry curvature and Weyl-node properties were evaluated from the Wannier tight-binding Hamiltonian using \textsc{WannierTools}~\cite{wu2018}. The Berry curvature of band $n$ was calculated using the Kubo-like formalism\cite{xiao2010berry}. Weyl nodes were identified from band crossings in the three-dimensional Brillouin zone, and their chirality was determined from the Berry-curvature flux through a closed surface enclosing each node. The Chern numbers were further evaluated from the Berry curvature over a closed two-dimensional manifold~\cite{yu2011,soluyanov2011},
\begin{equation}
	C=
	\frac{1}{2\pi}
	\int_{S}
	\Omega(k)\,d^{2}k,
	\label{eq:chern}
\end{equation}
and independently cross-checked using \textsc{Z2Pack}~\cite{gresch2017}. The intrinsic anomalous Hall conductivity (AHC) was calculated using \textsc{WannierBerri}~\cite{tsirkin2021} from the Wannier tight-binding Hamiltonian within the Berry-phase formulation of the anomalous Hall effect~\cite{nagaosa2010,xiao2010berry}.\\
The interatomic magnetic exchange parameters were extracted from the Wannier Hamiltonian using \textsc{TB2J}~\cite{He2021TB2J} package, based on the magnetic force theorem and the Green's-function formulation of Liechtenstein \textit{et al.}~\cite{Liechtenstein1987}. Magnon dispersions were calculated within linear spin-wave theory (LSWT) using the \textsc{SpinW} package~\cite{toth2015}, with the extracted exchange parameters provided as input. The spin operators were expanded around the $G$-type antiferromagnetic ground state using the Holstein--Primakoff transformation~\cite{holstein1940},
\begin{equation}
	S_i^z=S-a_i^\dagger a_i,\qquad
	S_i^+\simeq\sqrt{2S}\,a_i,\qquad
	S_i^-\simeq\sqrt{2S}\,a_i^\dagger,
	\label{eq:hp}
\end{equation}
which yields a quadratic bosonic Hamiltonian after Fourier transformation. The magnon energies were obtained through para-unitary diagonalization following the Colpa formalism~\cite{colpa1978}. 
\\
The magnetic ordering temperatures were obtained from classical
Metropolis Monte Carlo simulations using the unit-vector Heisenberg
Hamiltonian defined in Eq.~\eqref{eq:tb2j_heisenberg}. The Co magnetic moments were represented by classical unit vectors
$\mathbf{e}_i$. The bulk system was modeled using periodic $L\times L\times L$ supercells of the magnetic unit cell with $L=8$, $10$, $12$, and $14$, enabling finite-size-scaling analysis. Configurations were sampled using the single-spin-flip Metropolis algorithm~\cite{Metropolis1953,Newman1999}, with trial spin orientations drawn uniformly on the unit sphere and accepted with probability
\begin{equation}
	P = \min\left[1,\exp\left(-\frac{\Delta E}{k_{\mathrm{B}}T}\right)\right],
	\label{eq:metropolis}
\end{equation}
where $\Delta E$ is the corresponding energy change. Each Metropolis sweep was followed by a microcanonical over-relaxation sweep~\cite{Creutz1987,Brown1987} to reduce critical slowing down. At each temperature, $10^{4}$ combined sweeps were used for equilibration, followed by $2.5\times10^{4}$ sweeps for averaging over ten independent runs. The temperature grid was centered around a coarse transition estimate guided by the mean-field critical temperature obtained from the Fourier-transformed exchange matrix. For the compensated magnetic order, the staggered magnetization was evaluated as
\begin{equation}
	m_s = \frac{1}{N}\left\langle
	\Bigl|\sum_i \sigma_i \mathbf{e}_i\Bigr|
	\right\rangle,
	\label{eq:ms}
\end{equation}
where $\sigma_i$ denotes the sublattice phase. The magnetic susceptibility was calculated as $\chi = \frac{N}{k_{\mathrm{B}}T}
(\langle m_s^2\rangle-\langle m_s\rangle^2)$, together with the specific heat from energy fluctuations and the fourth-order Binder cumulant~\cite{Binder1981,Binder2010}
\begin{equation}
	U_L = 1 - \frac{\langle m_s^4\rangle}
	{3\langle m_s^2\rangle^{2}}.
	\label{eq:binder}
\end{equation}
 The N'eel temperature $T_N$ was determined primarily from the size-independent crossing of $U_L(T)$ and supported by the susceptibility peak. Its uncertainty was estimated from the spread of Binder-cumulant crossings among different system sizes. As an independent check, atomistic spin-dynamics simulations were performed using UppASD~\cite{skubic2008,Eriksson2017} with the same first-principles exchange parameters. The resulting transition temperatures, determined from the sublattice magnetization and specific-heat peak, agreed with the Monte Carlo Binder-crossing values within statistical uncertainty.
\section{Results and Discussion}
\subsection{Crystal and Magnetic Structure}\begin{figure}
	\centering
	\includegraphics[width=1\linewidth]{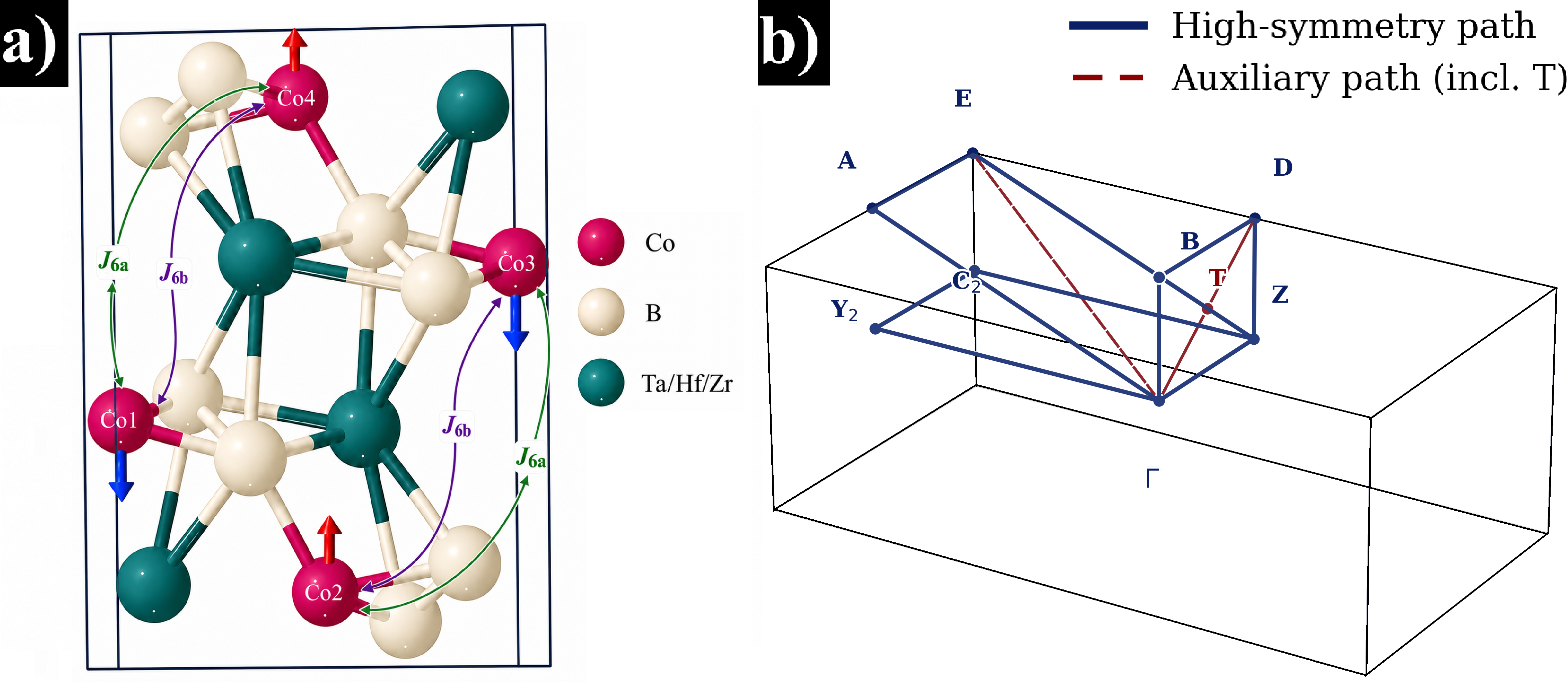}
	\caption{(a) Crystal structure of the $X$CoB$_2$ ($X =$ Ta, Zr, Hf) unit cell,
		showing the four Co sublattices (Co1--Co4, magenta), the boron (B) atoms, and the transition-metal sites ($X =$ Ta/Zr/Hf). Red and
		blue arrows on the Co atoms indicate the compensated collinear
		antiferromagnetic order (up/down sublattices). Curved arrows mark the
		dominant longer-range Co--Co exchange pathways $J_{6a}$ (green) and
		$J_{6b}$ (purple). (b) Corresponding Brillouin zone
		with the high-symmetry $k$-path used for the electronic and magnonic
		dispersions (solid blue, through $\Gamma$, A, E, C$_2$, Y$_2$, B, D, Z) and
		an auxiliary path including the T point (dashed red).}
	\label{fig:crystalbz}
\end{figure}
TaCoB$_2$, ZrCoB$_2$, and HfCoB$_2$ crystallize in the orthorhombic \textit{Pnma} space group (No.~62), belonging to the centrosymmetric point group \textit{mmm}. In all three compounds, the constituent atoms occupy the crystallographic $4c$ Wyckoff sites of the \textit{Pnma} structure. This structure type is well established experimentally, with TaCoB$_2$, together with the isostructural NbCoB$_2$, NbNiB$_2$, and TaNiB$_2$, having been synthesized and structurally characterized~\cite{steurer1978,Wind2014}. This experimentally established family of XYB$_2$ compounds includes combinations of group-5 and late $3d$ transition metals (Co, Ni). This indicates that the XYB$_2$ stoichiometry can form stable crystalline phases, supporting ZrCoB$_2$ and HfCoB$_2$ as plausible targets for synthesis. The optimized lattice parameters are summarized in Table~\ref{tab:lattice}, together with the available experimental values for TaCoB$_2$. The calculated lattice parameters of TaCoB$_2$ agree well with experiment, with deviations below $1.5\%$ along the $a$, $b$, and $c$ crystallographic axes. This agreement validates the optimization method and supports the predicted lattice parameters of ZrCoB$_2$ and HfCoB$_2$, for which experimental data are not yet available.
 \begin{table}[ht]
	\centering
	\caption{Optimized lattice parameters of TaCoB$_2$, ZrCoB$_2$, and HfCoB$_2$. Available experimental values are included for comparison. All lattice parameters are given in \AA.}
	\label{tab:lattice}
	\begin{tabular}{lccccccc}
		\toprule
		& \multicolumn{3}{c}{This work} & \multicolumn{3}{c}{Experiment} & \multirow{2}{*}{Reference} \\
		\cmidrule(lr){2-4}\cmidrule(lr){5-7}
		Compound & $a$ & $b$ & $c$ & $a$ & $b$ & $c$ & \\
		\midrule
		TaCoB$_2$ & 3.161 & 5.946 & 8.179 & 3.115 & 6.007 & 8.191 & \cite{Wind2014} \\
		ZrCoB$_2$ & 3.156 & 6.243 & 8.546 & -- & -- & -- & -- \\
		HfCoB$_2$ & 3.135 & 6.201 & 8.470 & -- & -- & -- & -- \\
		\bottomrule
	\end{tabular}
\end{table}
\\Figure \ref{fig:crystalbz}(a) shows the optimized crystal structure together with the magnetic ground-state configuration, while Figure \ref{fig:crystalbz}(b) presents the corresponding first Brillouin zone, including the high-symmetry points and the auxiliary (non-high-symmetry) momentum paths used in the electronic-structure calculations. The inclusion of these additional k-paths allows a more detailed investigation of spin-splitting phenomena beyond the conventional high-symmetry directions.\begin{figure}
\centering
\includegraphics[width=1\linewidth]{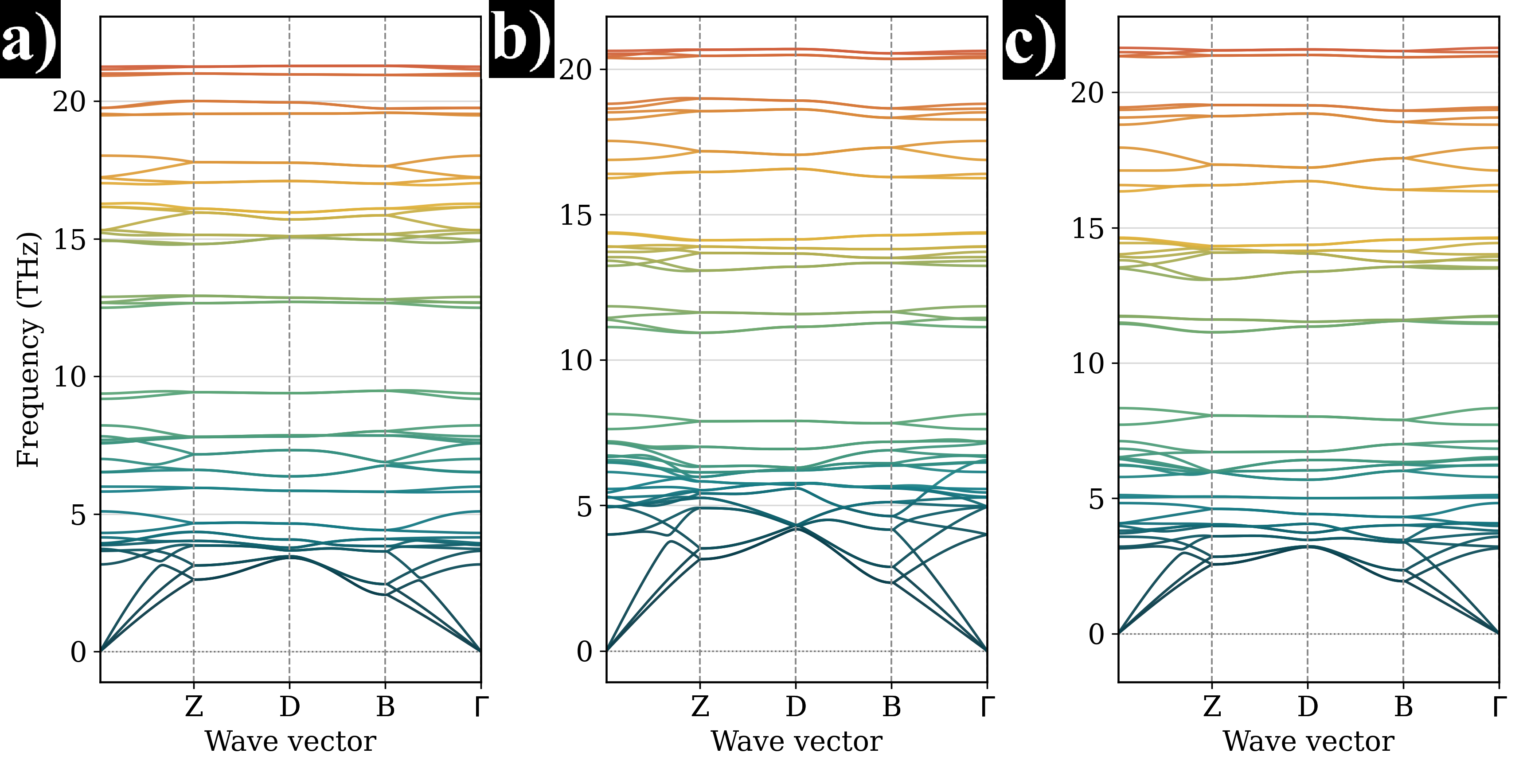}
\caption{Phonon dispersion relations of (a) TaCoB$_2$, (b) ZrCoB$_2$, and
	(c) HfCoB$_2$ along the $\Gamma$--Z--D--B--$\Gamma$ path, with branches
	colored by frequency. The absence of imaginary (negative) frequencies
	throughout the Brillouin zone confirms the dynamical stability of all three
	compounds in the assumed structure.}
\label{fig:fig-2}
\end{figure}\begin{table}[ht]
\centering
\small
\setlength{\tabcolsep}{3pt}
\caption{NM and AFM total energies, their difference
	$\Delta E = E_{\mathrm{NM}}-E_{\mathrm{AFM}}$, the Co local moment
	$m_{\mathrm{Co}}$, and the ground state (GS) in the AFM configuration
	(all energies per f.u.).}
\label{tab-groundstate}
\begin{tabular}{lccccc}
	\toprule
	Compound & $E_{\mathrm{NM}}$ & $E_{\mathrm{AFM}}$ & $\Delta E$ & $m_{\mathrm{Co}}$ & GS \\
	& (eV) & (eV) & (meV) & ($\mu_B$) & \\
	\midrule
	TaCoB$_2$ & $-32.0861$ & $-32.1366$ & $50.46$  & $1.16$ & AFM \\
	ZrCoB$_2$ & $-28.9212$ & $-29.2751$ & $353.93$ & $1.35$ & AFM \\
	HfCoB$_2$ & $-30.4356$ & $-30.7728$ & $337.23$ & $1.31$ & AFM \\
	\bottomrule
\end{tabular}
\end{table}
\begin{figure*}
	\centering
	\includegraphics[width=1\linewidth]{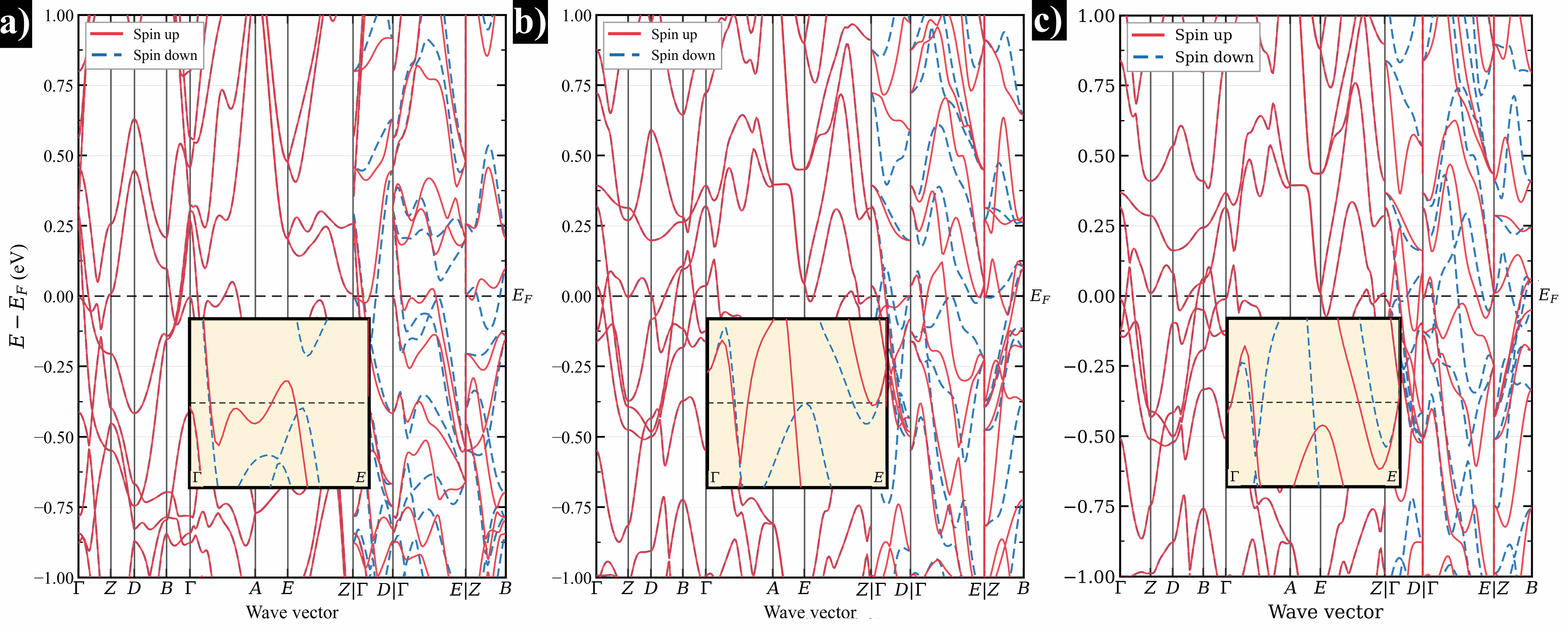}
	\caption{Spin-resolved electronic band structure of (a) TaCoB$_2$, (b) ZrCoB$_2$, 
		and (c) HfCoB$_2$ along selected high-symmetry and auxiliary 
		non-high-symmetry paths, with spin-up (red solid) and spin-down (blue 
		dashed) bands plotted relative to the Fermi level $E_F$ (horizontal dashed 
		line). The inset in each panel magnifies the $\Gamma$--$E$ segment near 
		$E_F$, highlighting the momentum-dependent spin splitting characteristic 
		of altermagnetism. The lifting of spin degeneracy along the auxiliary 
		paths, together with the symmetry-enforced degeneracy along specific 
		high-symmetry lines, is the hallmark of altermagnetic spin splitting. The 
		finite density of states crossing $E_F$ confirms the metallic character of 
		all three compounds.}
	\label{fig:fig-3}
\end{figure*}
The Co moments adopt a collinear G-type antiferromagnetic (AFM) arrangement. In
this configuration, each Co magnetic moment is antiparallel to all of its
nearest-neighbor Co moments, resulting in a compensated antiferromagnetic state
with zero net magnetization per unit cell. To assess the relative stability of the magnetic phases, self-consistent total-energy calculations were performed for the G-type AFM and nonmagnetic (NM) configurations of all three compounds. Ferromagnetic (FM) configurations were also examined, but no stable self-consistent FM solution could be obtained, as the FM initialization relaxed to the nonmagnetic state. The calculated total energies, energy differences, and Co local magnetic moments are summarized in Table~\ref{tab-groundstate}. The
G-type AFM configuration is energetically favored over the NM state for all
three compounds, with energy differences of $\Delta E = E_{\mathrm{NM}} -
E_{\mathrm{AFM}} = 50.46$, $353.93$, and $337.23$~meV/f.u.\ for TaCoB$_2$,
ZrCoB$_2$, and HfCoB$_2$, respectively. In the AFM state, the Co atoms carry
local magnetic moments of $1.16$, $1.35$, and $1.31~\mu_B$ for TaCoB$_2$,
ZrCoB$_2$, and HfCoB$_2$, respectively, aligned antiparallel on neighboring
sites in accordance with the G-type ordering. These results confirm the G-type
AFM configuration as the magnetic ground state of all three compounds.\subsection{Dynamic Stability} The dynamical stability of TaCoB$_2$, ZrCoB$_2$, and HfCoB$_2$ was investigated by calculating their phonon dispersion spectra using density-functional perturbation theory (DFPT). The phonon dispersions along the high-symmetry path $\Gamma\!-\!Z\!-\!D\!-\!B\!-\!\Gamma$ are presented in Figure~\ref{fig:fig-2}. Figure~\ref{fig:fig-2}(a) shows the phonon dispersion of TaCoB$_2$, while Figures~\ref{fig:fig-2}(b) and \ref{fig:fig-2}(c) show the corresponding dispersions for ZrCoB$_2$ and HfCoB$_2$, respectively. In all three compounds, no imaginary phonon frequencies are observed throughout the Brillouin zone, confirming their dynamical stability. The three acoustic branches approach zero frequency at the $\Gamma$ point, as required by translational invariance, while the remaining branches correspond to optical phonon modes. Since TaCoB$_2$ has already been synthesized experimentally~\cite{steurer1978,Wind2014}, the predicted dynamical stability of the isostructural ZrCoB$_2$ and HfCoB$_2$ suggests that they are also promising candidates for experimental synthesis.
\subsection{Symmetry and spin-splitting analysis}
Symmetry provides the fundamental framework for understanding the electronic structure of magnetic crystals. The crystallographic space group, together with the magnetic ordering, determines the symmetry operations that govern the electronic band structure and whether spin degeneracy is preserved or lifted. In this section, we briefly discuss the symmetry properties of the orthorhombic $Pnma$ (No.~62) crystal structure and their implications for the electronic structures of the investigated compounds. The symmetry of the magnetically ordered phase is described by the corresponding magnetic space group (MSG), which incorporates both the crystallographic symmetry and the magnetic configuration. For the G-type antiferromagnetic order considered here, the four Co atoms occupying the $4c$ Wyckoff positions form two magnetic sublattices with antiparallel spin moments. The resulting magnetic symmetry is described by the magnetic space group $Pnm'a'$ (BNS No.~62.447). Within this MSG, the two magnetic sublattices are connected by nonsymmorphic glide-mirror operations combined with time-reversal symmetry, $\mathcal{T}$. This satisfies the symmetry criterion for altermagnetism established by \v{S}mejkal \textit{et al.}~\cite{Smejkal2022a,Smejkal2022b}. The corresponding nonrelativistic magnetic space group is the type-III MSG $Pnm'a'$ (BNS No.~62.447 \cite{gallego2016magndata,belov1957shubnikov}), which breaks both the combined inversion--time-reversal symmetry, $\mathcal{PT}$, and the spin-rotation--translation symmetry, $U\tau$~\cite{Rooj2025}. As a result, opposite-spin electronic states are no longer constrained to remain degenerate throughout the Brillouin zone, allowing momentum-dependent spin splitting to develop away from symmetry-protected regions.\\From our symmetry analysis of the G-type antiferromagnetic configuration, the eight spatial operations of the parent space group $Pnma$ are partitioned into two sets, defining the type-III magnetic space group $Pnm'a'$. Four of these remain ordinary unitary operations,
\begin{equation}
	G_{U} = \bigl\{1\,|\,\mathbf{0}\bigr\},\;
	\bigl\{2_{100}\,|\,\tfrac{1}{2}\tfrac{1}{2}\tfrac{1}{2}\bigr\},\;
	\bigl\{\bar{1}\,|\,\mathbf{0}\bigr\},\;
	\bigl\{m_{100}\,|\,\tfrac{1}{2}\tfrac{1}{2}\tfrac{1}{2}\bigr\},
	\label{eq:GU}
\end{equation}
each of which maps every spin-up site onto a spin-up site and every spin-down site onto a spin-down site, leaving the magnetic configuration invariant
($\uparrow\rightarrow\uparrow$, $\downarrow\rightarrow\downarrow$). The remaining four operations interchange the two magnetic sublattices
($\uparrow\leftrightarrow\downarrow$) and therefore are not symmetries by themselves. They become symmetries only when combined with time-reversal symmetry, $\mathcal{T}$, which reverses all spins and restores the original magnetic configuration. These constitute the antiunitary operations,
\begin{align}
	G_{AU} = 
	&\mathcal{T}\bigl\{2_{001}\,|\,\tfrac{1}{2}0\tfrac{1}{2}\bigr\},\;
	\mathcal{T}\bigl\{2_{010}\,|\,0\tfrac{1}{2}0\bigr\},\notag\\
	&\mathcal{T}\bigl\{m_{001}\,|\,\tfrac{1}{2}0\tfrac{1}{2}\bigr\},\;
	\mathcal{T}\bigl\{m_{010}\,|\,0\tfrac{1}{2}0\bigr\}.
	\label{eq:GAU}
\end{align}
An antiunitary operation $A \in G_{AU}$ enforces spin degeneracy
$E_{\uparrow}(\mathbf{k}) = E_{\downarrow}(\mathbf{k})$ at a wave vector
$\mathbf{k}$ if and only if $A$ maps $\mathbf{k}$ onto itself (modulo a
reciprocal-lattice vector). For $\{R\,|\,\mathbf{t}\}\cdot\mathcal{T}$, the
action on $\mathbf{k}$ is
\begin{equation}
	\mathbf{k} \;\xrightarrow{\{R|\mathbf{t}\}\cdot\mathcal{T}}\; -R\,\mathbf{k}
	\quad (\mathrm{mod\ recip.\ lattice}).
	\label{eq:kaction}
\end{equation}
Applying Eq.~\eqref{eq:kaction} to the four antiunitary operations in
Eq.~\eqref{eq:GAU} shows that $\mathcal{T}\{2_{001}|\frac{1}{2}0\frac{1}{2}\}$
leaves the entire $k_{z}=0$ plane invariant, and
$\mathcal{T}\{2_{010}|0\frac{1}{2}0\}$ leaves the entire $k_{y}=0$ plane
invariant. No element of $G_{AU}$ self-maps a generic point in the $k_{x}=0$
plane; that plane is fixed only by the unitary mirror
$\{m_{100}|\frac{1}{2}\frac{1}{2}\frac{1}{2}\}$, which does not exchange the
sublattices and therefore does not enforce spin degeneracy. The protected and
split regions of the Brillouin zone are therefore:
\begin{equation}
	E_{\uparrow}(\mathbf{k}) = E_{\downarrow}(\mathbf{k})
	\iff k_{y} \in \bigl\{0,\tfrac{1}{2}\bigr\}
	\;\text{ or }\;
	k_{z} \in \bigl\{0,\tfrac{1}{2}\bigr\},
	\label{eq:master}
\end{equation}
while spin splitting is symmetry-allowed wherever $k_{y} \notin \{0,\frac{1}{2}\}$
\emph{and} $k_{z} \notin \{0,\frac{1}{2}\}$, with the $k_{x}=0$ plane being
the principal altermagnetic plane.\\ 
The transformation of the splitting function
$\Delta(\mathbf{k})\equiv E_{\uparrow}(\mathbf{k})-E_{\downarrow}(\mathbf{k})$
under $G_{U}$ and $G_{AU}$ fixes its momentum parity. A unitary operation
carries $\Delta$ to its image $\mathbf{k}$-point with the same sign, since it
preserves the spin label, whereas an antiunitary operation flips the spin
label and therefore relates $\Delta(\mathbf{k})$ to $-\Delta(-R\mathbf{k})$.
Applying this to the generators, $\mathcal{T}\{2_{010}|0\tfrac{1}{2}0\}$ maps
$\mathbf{k}\!\to\!(k_x,-k_y,k_z)$ and so makes $\Delta$ odd in $k_{y}$;
$\mathcal{T}\{2_{001}|\tfrac{1}{2}0\tfrac{1}{2}\}$ maps
$\mathbf{k}\!\to\!(k_x,k_y,-k_z)$ and makes it odd in $k_{z}$; while the
unitary mirror $\{m_{100}|\tfrac{1}{2}\tfrac{1}{2}\tfrac{1}{2}\}$ maps
$\mathbf{k}\!\to\!(-k_x,k_y,k_z)$ with no sign change, keeping $\Delta$ even
in $k_{x}$. The lowest-order symmetry-allowed term in
the $\mathbf{k}\cdot\mathbf{p}$ expansion of the spin splitting about $\Gamma$
is~\cite{Rooj2025}
\begin{equation}
	\Delta(\mathbf{k}) \;\propto\; k_{y}k_{z},
	\label{eq:dyz}
\end{equation}
identifying the spin splitting as a $d_{yz}$-wave whose nodes lie on the
$k_{y}=0$ and $k_{z}=0$ planes. Spin degeneracy is thus enforced wherever
$k_{y}\in\{0,\tfrac{1}{2}\}$ or $k_{z}\in\{0,\tfrac{1}{2}\}$
[Eq.~\eqref{eq:master}], and splitting is symmetry-allowed only in the
interior of the $k_{x}=0$ face. Table~\ref{tab:paths} classifies the
high-symmetry paths used in this work on this basis: degeneracy is lifted
only along $\Gamma\!-\!D$, $\Gamma\!-\!T$, $\Gamma\!-\!E$, and $Z\!-\!B$.
\begin{table}[t]
	\caption{\label{tab:paths}%
		Predicted spin degeneracy and spin splitting along the high-symmetry and auxiliary paths considered in this work.}
	\begin{ruledtabular}
		\begin{tabular}{lcc}
			Path & $(k_{x},k_{y},k_{z})$ & Symmetry result \\
			\hline
			$\Gamma\!\to\!Z$  & $(0,t,0)$                 & Degenerate \\
			$Z\!\to\!D$       & $(0,\tfrac{1}{2},t)$      & Degenerate \\
			$D\!\to\!B$       & $(0,\tfrac{1}{2}{-}t,\tfrac{1}{2})$ & Degenerate \\
			$B\!\to\!\Gamma$  & $(0,0,t)$                 & Degenerate \\
			$\Gamma\!\to\!A$  & $(-t,0,t)$                & Degenerate \\
			$A\!\to\!E$       & $(-\tfrac{1}{2},t,\tfrac{1}{2})$ & Degenerate \\
			$E\!\to\!Z$       & $(-\tfrac{1}{2}{+}t,\tfrac{1}{2},\tfrac{1}{2}{-}t)$ & Degenerate \\
			$\Gamma\!\to\!D$  & $(0,t,t)$                 & Split \\
			$\Gamma\!\to\!E$  & $(-t,t,t)$                & Split \\
			$Z\!\to\!B$       & $(0,\tfrac{1}{2}{-}t,t)$  & Split \\
		\end{tabular}
	\end{ruledtabular}
\end{table}
\subsection{Electronic band structure}
\begin{figure*}
	\centering
	\includegraphics[width=1\linewidth]{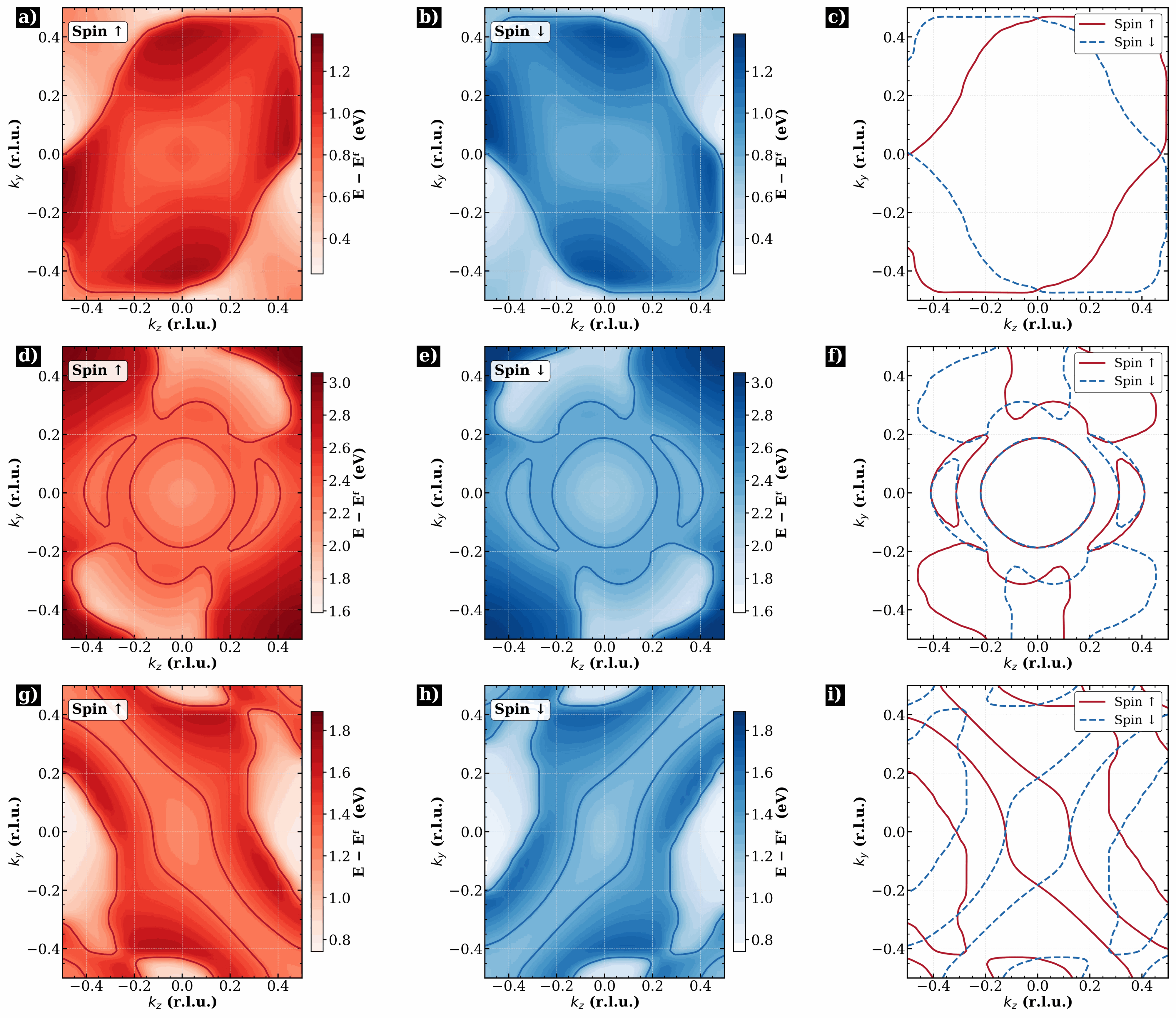}
	\caption{Spin-polarized iso-energy maps on the $k_x = 0$ plane for TaCoB$_2$
		(top row, a--c), ZrCoB$_2$ (middle row, d--f), and HfCoB$_2$ (bottom row,
		g--i). For each compound the iso-energy surface is evaluated at the energy
		of maximum spin splitting on this plane, namely $E - E_F = +0.80$~eV for
		TaCoB$_2$, $+2.32$~eV for ZrCoB$_2$, and $+1.32$~eV for HfCoB$_2$. The left
		(a,d,g) and middle (b,e,h) columns show the spin-up and spin-down channels,
		respectively, and the right column (c,f,i) overlays the corresponding
		spin-up (solid red) and spin-down (dashed blue) contours. The two spin
		channels display distinctly different topologies, with their contours
		rotated by $90^\circ$ relative to one another in the $k_y$--$k_z$ plane,
		consistent with the $d_{yz}$-wave form factor
		$\Delta(\mathbf{k}) \propto k_y k_z$. The nodal lines of the splitting
		function, along which the two spin channels coincide, lie on the $k_y = 0$
		and $k_z = 0$ axes in all three materials. This spin-momentum locking
		identifies the compounds as $d_{yz}$-wave altermagnets within the magnetic
		space group $Pnm'a'$.}
	\label{fig:fig-4}
\end{figure*}
The spin-polarized electronic band structures of TaCoB$_2$, ZrCoB$_2$, and HfCoB$_2$ are shown in Figure~\ref{fig:fig-3}(a)--\ref{fig:fig-3}(c), respectively. The calculations were performed along a high-symmetry path that passes through both the symmetry-protected degenerate regions and the spin-split regions of the Brillouin zone identified in Sec. B. All three compounds are metallic, with multiple bands crossing the Fermi level. A prominent feature of the band structures is the pronounced directional dependence of the spin-resolved electronic states: along specific high-symmetry lines the spin-up (red) and spin-down (dashed blue) bands remain exactly degenerate, whereas along other directions this degeneracy is lifted, resulting in pronounced spin splitting both near and away from the Fermi level. These first-principles results faithfully reproduce the symmetry-protected spin-degenerate and spin-split regions predicted by the group-theoretical analysis summarized in Table~\ref{tab:paths}, and are fully consistent with the $d_{yz}$-wave spin-splitting form factor derived in Sec.~B.\\
Along the high-symmetry paths $\Gamma$-$Z$-$D$-$B$-$\Gamma$ and $\Gamma$-$A$-$E$-$Z$, the two spin channels remain perfectly degenerate throughout all three materials, as is clearly visible in Figs.~\ref{fig:fig-3}(a)--\ref{fig:fig-3}(c). As established in Sec.~B, these paths lie on the $k_y \in \{0,\frac{1}{2}\}$ or $k_z \in \{0,\frac{1}{2}\}$ planes, where the antiunitary operations $\mathcal{T}\{2_{001}|\frac{1}{2}0\frac{1}{2}\}$ and $\mathcal{T}\{2_{010}|0\frac{1}{2}0\}$ self-map the relevant $\mathbf{k}$ points and thus enforce $E_\uparrow(\mathbf{k}) = E_\downarrow(\mathbf{k})$ [Eq.~\eqref{eq:master}]. This spin degeneracy is therefore not a coincidental near-degeneracy but a symmetry-enforced consequence of the MSG $Pnm'a'$ (BNS No.~62.447), as established by the preceding symmetry analysis. The full degeneracy of the $k_x$--$k_z$ and $k_x$--$k_y$ planes, where all bands remain doubly degenerate across the entire face of the Brillouin zone, is confirmed in Supplementary Figs.~S-1 and S-2, respectively.
\\In stark contrast, the three auxiliary paths $\Gamma\!\to\!D$ (along $(0,t,t)$), $Z\!\to\!B$ (along $(0,\frac{1}{2}-t,t)$), and $\Gamma\!\to\!E$ (along $(-t,t,t)$) simultaneously have $k_y \notin \{0,\frac{1}{2}\}$ and $k_z \notin \{0,\frac{1}{2}\}$, placing them in the interior of the $k_x = 0$ face where no element of $G_{AU}$ self-maps a generic $\mathbf{k}$ point. Along these paths, large spin splittings are clearly visible in 
Figure~\ref{fig:fig-3}(a) for TaCoB$_2$, Figure~\ref{fig:fig-3}(b) for 
ZrCoB$_2$, and Figure~\ref{fig:fig-3}(c) for HfCoB$_2$, with the spin-up 
and spin-down manifolds separating by significant fractions of an eV near 
the Fermi level. To quantify the magnitude of the altermagnetic splitting 
on the $k_x = 0$ plane, we computed the momentum-averaged spin splitting 
$\langle\Delta\rangle$ both at the Fermi level and at the energy where it 
is maximised. At the Fermi level, the averaged splitting reaches 88.38~meV 
for TaCoB$_2$, 52.34~meV for ZrCoB$_2$, and 70.61~meV for HfCoB$_2$. The maximum splitting for TaCoB$_2$ is 227.60~meV, attained at $E - E_F = +0.80$~eV above the Fermi level. For ZrCoB$_2$, the maximum splitting reaches 200.73~meV at $E - E_F = +2.32$~eV, while for HfCoB$_2$ it reaches 216.81~meV at $E - E_F = +1.32$~eV. The large magnitude of the spin splitting, exceeding 200~meV in all three compounds, suggests that these materials are promising candidates for altermagnetic spintronic applications, including spin-current generation~\cite{GonzalezHernandez2021,Bai2023PRL}, spin filtering~\cite{Smejkal2022TMR}, and high-density magnetic memories and terahertz nano-oscillators~\cite{Song2025NRM,Jungwirth2025NP,Bai2024AFM}.\begin{figure*}
	\centering
	\includegraphics[width=1\linewidth]{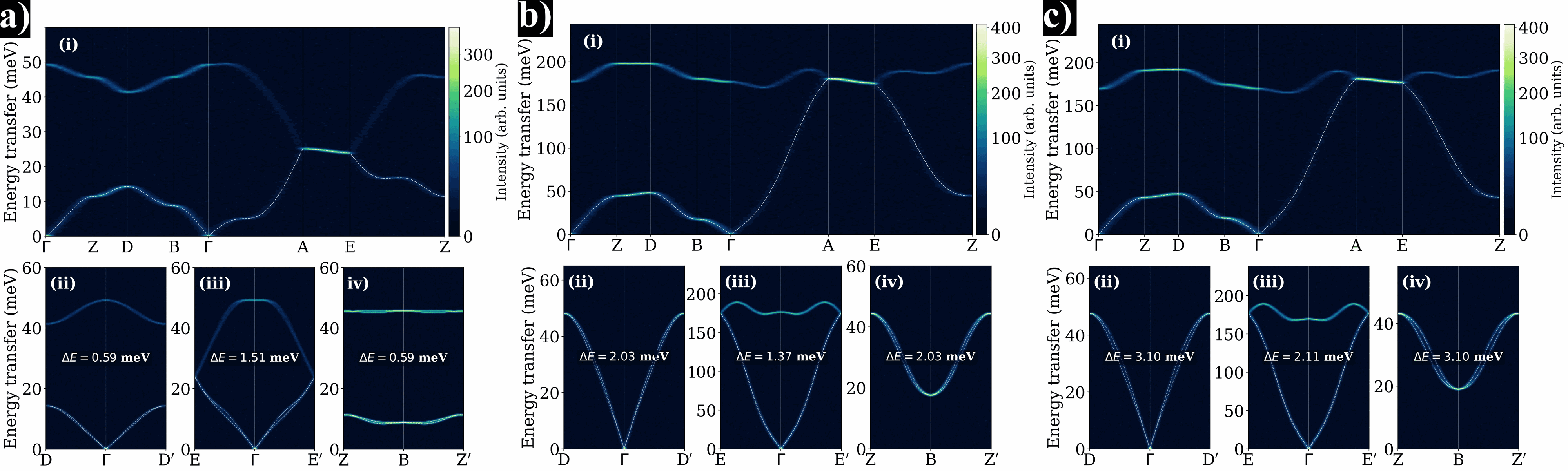}
	\caption{Simulated INS intensity maps of the magnon dispersion for 
		(a)~TaCoB$_2$, (b)~ZrCoB$_2$, and (c)~HfCoB$_2$, computed within LSWT using 
		TB2J exchange parameters. Panel~(i) shows the full spectrum along 
		$\Gamma$--Z--D--B--$\Gamma$--A--E--Z; panels~(ii)--(iv) show the acoustic 
		branches along the auxiliary paths D--$\Gamma$--D$'$, E--$\Gamma$--E$'$, and 
		Z--B--Z$'$, where the two-fold magnon degeneracy is lifted --- a spin-wave 
		signature of altermagnetism. White dashed lines are the calculated dispersion 
		curves.}
	\label{fig:fig-5}
\end{figure*}
\\
The altermagnetic character is further supported by the spin-polarized iso-energy maps on the $k_x = 0$ plane, shown in Figure~\ref{fig:fig-4}(a)-(i). The three rows correspond to TaCoB$_2$, ZrCoB$_2$, and HfCoB$_2$, and for each material the iso-energy surface is evaluated at the energy of maximum spin splitting identified above, so that the spin contrast is rendered as prominently as possible. The two spin channels display distinctly different topologies on the $k_x = 0$ face, with the spin-up and spin-down iso-energy contours rotated relative to each other in a pattern fully consistent with the $d_{yz}$-wave form factor $\Delta(\mathbf{k}) \propto k_y k_z$ [Eq.~\eqref{eq:dyz}]. The nodal lines of the splitting function, where the contours of the two spin channels must cross or coincide, lie precisely on the $k_y = 0$ and $k_z = 0$ axes in all three materials, in full accord with Eq.~\eqref{eq:master}. The overlaid contours clearly show this spin-momentum locking. The two contours are related by mirror symmetry about either nodal axis, which exchanges the spin states and reverses the sign of the splitting, consistent with the antiunitary generators in Eq.~\eqref{eq:GAU}. This four-fold rotational asymmetry between the spin-up and spin-down sheets, related by a $90^\circ$ rotation in the $k_y$--$k_z$ plane, is the momentum-space signature that unambiguously classifies these compounds as $d_{yz}$-wave altermagnets within the MSG $Pnm'a'$.
\subsection{Microscopic Origin of Chiral Magnon Splitting}
To identify the exchange interactions responsible for the chiral
magnon splitting in XCoB$_2$ ($X=\mathrm{Ta}$, Zr, and Hf), we map the
magnetic energy onto a classical Heisenberg model,
\begin{equation}
	\mathcal{H}_{\mathrm{ex}}
	= -\sum_{i\neq j} J_{ij}\,\hat{\mathbf{S}}_i\cdot\hat{\mathbf{S}}_j,
	\label{eq:tb2j_heisenberg}
\end{equation}
where $\hat{\mathbf{S}}_i$ is a unit vector along the magnetic moment of
the $i$th Co atom, $J_{ij}$ is the exchange coupling between
Co sites $i$ and $j$, and the summation runs over all ordered pairs, so
that each interacting pair is counted twice. In this convention,
$J_{ij}>0$ favours ferromagnetic alignment, whereas $J_{ij}<0$ favours
antiferromagnetic alignment. The couplings were obtained from the
Wannier-based magnetic force theorem as implemented in
\textsc{TB2J}~\cite{He2021TB2J}, using the maximally localised Wannier
functions. The cutoff $R_c$ sets the maximum Co to Co distance out to which pairs are included, and we evaluate all interactions up to $R_c=10$~\AA. Starting from any Co atom, we order its neighbouring Co atoms by their Co to Co
	distance. The closest Co atoms are the first nearest neighbours, the next group
	out are the second nearest neighbours, and so on. Co atoms at the same distance
	are related by the crystal symmetry and share the same coupling, so we label $J_n$ according to their neighbour order ($J_1$, $J_2$, $\ldots$, $J_6$). Thus, each $J_n$ in Table~\ref{tab:exchange} is
	the coupling that describes the $n$th nearest neighbours of Co. The
	sixth nearest neighbour is the one exception, as it consists of two sets of Co
	pairs at nearly the same distance that are not related by symmetry, which we
	label $J_{6a}$ and $J_{6b}$. As shown in Figure~S-3, the exchange interactions alternate in  sign with distance rather than decaying monotonically. A short-range exchange description is therefore insufficient.\\
Figure~\ref{fig:fig-5} presents the simulated inelastic neutron scattering (INS) 
spectra of (a) TaCoB$_2$, (b) ZrCoB$_2$, and (c) HfCoB$_2$, obtained from 
linear spin-wave theory using the exchange parameters in Table~\ref{tab:exchange}.\begin{table*}[t]
	\centering
	\caption{Heisenberg exchange parameters $J_{ij}$ (in meV) for
		TaCoB$_2$, ZrCoB$_2$, and HfCoB$_2$. $J>0$ and $J<0$ denote ferromagnetic
		and antiferromagnetic interactions, respectively. Each interaction is labelled intra- or inter-sublattice
			depending on whether it links Co atoms of the same spin or opposite spin.
			The value $d$ is the distance between the two Co atoms, not the length of
			the path that connects them through the neighbouring atoms. Each $d$ is given as a range because the same neighbour occurs at a slightly
			different Co--Co distance in TaCoB$_2$, ZrCoB$_2$, and HfCoB$_2$. The sixth nearest neighbour splits into two inequivalent pathways, $J_{6a}$ and $J_{6b}$, whose
			difference $\Delta J_{\mathrm{AM}}=|J_{6b}-J_{6a}|$ characterizes the
			altermagnetic exchange inequivalence.}
	\label{tab:exchange}
	\begin{ruledtabular}
		\begin{tabular}{lccccc}
			& Sublattice & $d$ (\AA)
			& TaCoB$_2$ & ZrCoB$_2$ & HfCoB$_2$ \\
			\colrule
			
			$J_{1}$  
			& intra & 2.43--2.44
			& $4.887$ & $18.692$ & $17.669$ \\
			
			$J_{2}$  
			& intra & 3.14--3.16
			& $-0.902$ & $9.007$ & $9.549$ \\
			
			$J_{3}$  
			& inter & 3.73--3.95
			& $-2.255$ & $-2.591$ & $-2.710$ \\
			
			$J_{4}$  
			& inter & 4.89--5.05
			& $-0.296$ & $-1.311$ & $-1.355$ \\
			
			$J_{5}$  
			& intra & 5.06--5.09
			& $-0.618$ & $0.746$ & $0.911$ \\
			
			$J_{6a}$  
			& inter & 5.25--5.50
			& $-0.591$ & $0.911$ & $0.950$ \\
			
			$J_{6b}$  
			& inter & 5.35--5.54
			& $0.422$ & $-0.887$ & $-0.966$ \\
			
			\colrule
			
			$\Delta J_{\mathrm{AM}}$
			& \multicolumn{2}{c}{$|J_{6b}-J_{6a}|$}
			& $1.014$ & $1.798$ & $1.916$ \\
			
		\end{tabular}
	\end{ruledtabular}
\end{table*} 
For each compound, the upper panel (i) shows the magnon dispersion along the high-symmetry path $\Gamma-Z-D-B-\Gamma-A-E-Z'$. The lower panels (ii)--(iv) show the simulated INS spectra along the non-high-symmetry directions $D-\Gamma-D'$, $E-\Gamma-E'$, and $Z-B-Z'$, respectively, where the chiral splitting becomes most pronounced. Along these directions, the two magnon branches of opposite chirality, become nondegenerate, while they remain degenerate along the symmetry-protected directions. The maximum splitting reaches $1.51$~meV in TaCoB$_2$ along $E-\Gamma-E'$, $2.03$~meV in ZrCoB$_2$ along $D-\Gamma-D'$, and $3.10$~meV in HfCoB$_2$ along $D-\Gamma-D'$. The momentum dependence of the splitting is consistent with the
altermagnetic symmetry discussed in Sec.~B. These results demonstrate
the presence of chiral magnon splitting in all three compounds and
motivate a detailed investigation of the microscopic exchange
interactions responsible for lifting the magnon degeneracy.\\
\begin{table}[t]
	\caption{\label{tab:splitting-comparison}%
		Chiral magnon splitting $\Delta\varepsilon_{\max}$ reported for
		altermagnets, together with the excitation energy $\varepsilon$ at
		which the maximum splitting occurs. INS denotes inelastic neutron
		scattering, LSWT linear spin-wave theory, and LR-TDDFT
		linear-response time-dependent density functional theory.}
	\begin{ruledtabular}
		\begin{tabular}{lccc}
			Material & Method & $\Delta\varepsilon_{\max}$ (meV) & $\varepsilon$ (meV) \\
			\colrule
			$\alpha$-MnTe~\cite{Liu2024MnTe}            & INS      & $\approx 2$  & 30-35      \\
			$\alpha$-Fe$_2$O$_3$~\cite{Hoyer2025Hematite} & LSWT   & $\approx 2$  &  $\sim 100$ \\
			$\alpha$-Fe$_2$O$_3$~\cite{Sun2025Hematite}   & INS    & $\approx 3$  & $\sim 100$           \\
			NaOsO$_3$~\cite{NaOsO3_2025}                & LSWT     & $\approx 5$  & $\sim 150$    \\
			CuF$_2$ ~\cite{Ho2026CuF2}          & LSWT     & $\approx 2$    & 10           \\
			\colrule
			TaCoB$_2$ (this work) & LSWT & $1.51$ & $25$ to $50$  \\
			ZrCoB$_2$ (this work) & LSWT & $2.03$ & $\lesssim 50$ \\
			HfCoB$_2$ (this work) & LSWT & $3.10$ & $\lesssim 50$ \\
		\end{tabular}
	\end{ruledtabular}
\end{table}The calculated splittings in XCoB$_2$ are comparable to those reported for other altermagnets, as summarized in Table~\ref{tab:splitting-comparison}. A splitting of approximately $2$~meV has been observed by INS in $\alpha$-MnTe~\cite{Liu2024MnTe}, while hematite shows values of about $2$--$3$~meV~\cite{Hoyer2025Hematite,Sun2025Hematite}. A larger splitting of approximately $5$~meV has been predicted for the optic branch of NaOsO$_3$~\cite{NaOsO3_2025}. Notably, the largest splittings in ZrCoB$_2$ and HfCoB$_2$ occur below approximately $50$~meV, while the entire magnon spectrum of TaCoB$_2$ lies below this energy. This contrasts with hematite and NaOsO$_3$, where the splitting occurs above $100$~meV~\cite{Sun2025Hematite,NaOsO3_2025}. Such a low excitation-energy scale is advantageous for INS because it can be accessed with lower incident neutron energies. Chiral magnon splitting can also support field-free spin Seebeck and spin Nernst responses~\cite{Cui2023,Weissenhofer2024}. Together with the strong directional dependence found here and the recently reported direction-dependent zero-field spin Seebeck response in LuFeO$_3$~\cite{Galindez2025}, these results suggest potential for anisotropic field-free magnon transport in XCoB$_2$.
\\
To determine the microscopic origin of the splitting, we examine the exchange parameters in Table~\ref{tab:exchange}. Among the first five nearest neighbours, $J_1$ and $J_2$ are the dominant interactions in ZrCoB$_2$ and HfCoB$_2$. The ferromagnetic $J_1$ values are $18.692$ and $17.669$~meV, respectively, while $J_2$ is also ferromagnetic at $9.007$ and $9.549$~meV. In contrast, $J_3$ and $J_4$ are antiferromagnetic in all three compounds. TaCoB$_2$ exhibits a weaker and modified exchange hierarchy, with $J_1=4.887$~meV, $J_2=-0.902$~meV, and $J_5=-0.618$~meV. These variations show that the $X$-site element strongly modifies the magnetic exchange network.
\\
We then perform a systematic exchange-cutoff analysis to determine how these interactions contribute to the magnon spectrum. When interactions are retained only through $J_5$, the spectra in Figure~S-4 retain the principal features of the magnon dispersion and the overall spin-wave energy scale, but the branches remain degenerate along the low-symmetry directions. Thus, $J_1$--$J_5$ establish the main magnetic background but are insufficient to generate chiral splitting. Upon including the sixth inter-sublattice nearest neighbour, a clear splitting emerges, as shown in Figure~\ref{fig:fig-5}(ii)--(iv). This identifies the sixth nearest neighbour as the first one required to lift the magnon degeneracy within the Heisenberg framework, while more distant interactions mainly provide quantitative corrections to the splitting magnitude. Importantly, the sixth nearest neighbour does not consist of a single magnetically equivalent interaction. The altermagnetic crystal symmetry divides it into two crystallographically inequivalent inter-sublattice exchange paths. These are denoted $J_{6a}$ and $J_{6b}$ in Table~\ref{tab:exchange} and are depicted in Figure~\ref{fig:crystalbz}(a). The two paths have similar but distinct Co--Co separations and different exchange parameters. Here, $J_{6a}$ and $J_{6b}$ denote the shorter and longer Co--Co paths, respectively. If the opposite-spin sublattices were related by translation or inversion, the two interactions would be equivalent and have identical exchange constants. In XCoB$_2$, however, the altermagnetic crystal symmetry allows them to be inequivalent~\cite{Smejkal2022a}, and their exchange couplings can therefore differ. We quantify this inequivalence by defining
\begin{equation}
	\Delta J_{\mathrm{AM}}
	=
	\left|J_{6b}-J_{6a}\right|,
	\label{eq:deltaJAM}
\end{equation}
which measures the long-range altermagnetic exchange inequivalence associated with the sixth-neighbor shell.\begin{figure*}
	\centering
	\includegraphics[width=1\linewidth]{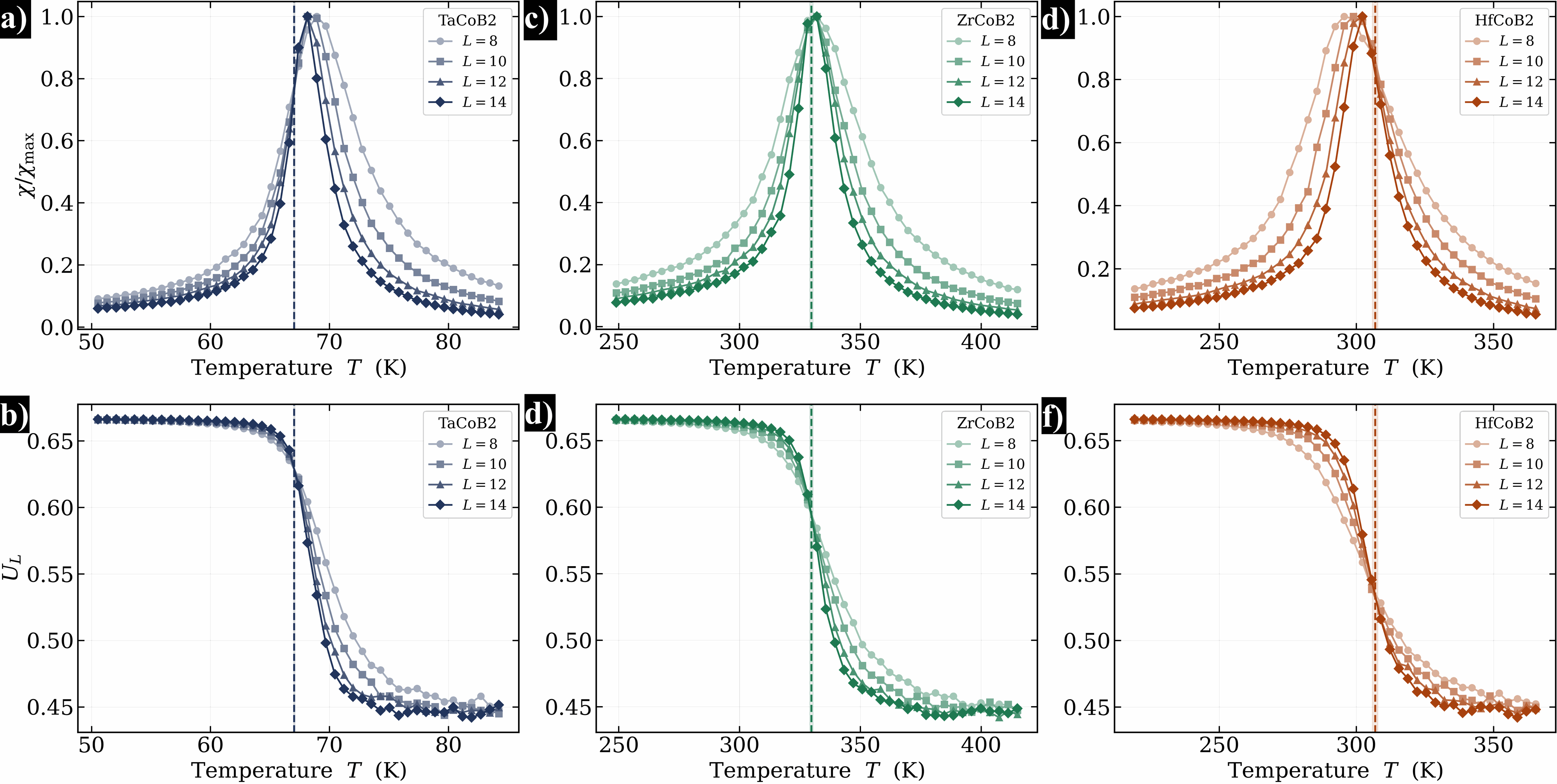}
	\caption{Finite-size Monte Carlo determination of the N\'eel temperature for
		(a,d)~TaCoB$_2$, (b,e)~HfCoB$_2$, and (c,f)~ZrCoB$_2$ using linear system
		sizes $L = 8$--$14$. Top row: normalized magnetic susceptibility
		$\chi/\chi_{\max}$. Bottom row: Binder cumulant $U_L$. Dashed vertical lines denote $T_N$, obtained from the
		size-independent crossing of $U_L$. The susceptibility maxima coincide with
		the cumulant crossings and grow with system size, confirming a genuine
		thermodynamic phase transition in each compound.}
	\label{fig:fig-6}
\end{figure*}\\A striking feature of Table~\ref{tab:exchange} is that $J_{6a}$ and $J_{6b}$ have opposite signs in all three compounds. In $\mathrm{TaCoB}2$, $J_{6a}=-0.591$~meV and $J_{6b}=+0.422$~meV, giving $\Delta J_{\mathrm{AM}}=1.014$~meV. The magnitude of the exchange inequivalence increases to $1.798$~meV in $\mathrm{ZrCoB}2$, where $J_{6a}=+0.911$~meV and $J_{6b}=-0.887$~meV, and reaches $1.916$~meV in $\mathrm{HfCoB}2$, with $J_{6a}=+0.950$~meV and $J_{6b}=-0.966$~meV. Thus, $\Delta J_{\mathrm{AM}}$ increases systematically from Ta to Zr to Hf. The maximum magnon splitting follows the same trend, increasing from $1.51$~meV in TaCoB$_2$ to $2.03$~meV in ZrCoB$_2$ and $3.10$~meV in HfCoB$_2$. This correlation supports the role of the sixth-neighbor exchange inequivalence in controlling the chiral magnon splitting, while more distant interactions can further modify its magnitude.\\
Our analysis shows that the exchange inequivalence between $J_{6a}$ and $J_{6b}$ originates primarily from their different microscopic exchange pathways rather than from the Co--Co separation itself. Both interactions connect opposite Co spin sublattices and have fractional bond vectors of the form $(\tfrac12,\tfrac12\pm\delta,\tfrac12)$, where $\delta$ denotes the displacement of the Co sublattice from the ideal $y=\tfrac12$ position. As a result, the corresponding Co--Co distances are very similar. For example, in HfCoB$_2$ they are $5.462$ and $5.493$~\AA. Despite this small difference in distance, the two interactions have different signs and magnitudes, indicating that their inequivalence is governed mainly by the distinct intermediate atomic configurations.
\\HfCoB$_2$ provides a representative example. Although the Co--Co separations for $J_{6a}$ and $J_{6b}$ are both approximately $5.4$~\AA, the exchange is mediated through different atomic pathways. For $J_{6a}$, the Co--B--B--Co pathway contains Co--B distances of $2.153$ and $2.199$~\AA\ and a B--B distance of $1.815$~\AA. In contrast, $J_{6b}$ follows a Co--B--Hf--Co pathway, with Co--B, B--Hf, and Hf--Co distances of $2.089$, $2.559$, and $2.773$~\AA, respectively. Thus, the nearly identical Co--Co separations involve distinct intermediate atoms and exchange pathways, consistent with the finite exchange inequivalence $\Delta J_{\mathrm{AM}}$. The same trend persists across the $X$CoB$_2$ series. The boron-bridged pathway is ferromagnetic, whereas the $X$-mediated pathway is antiferromagnetic, with similar boron-bridge geometry characterized by Co--B distances of $2.05$--$2.24$~\AA\ and B--B distances of $1.815$--$1.861$~\AA. In $\mathrm{ZrCoB}_2$ and $\mathrm{HfCoB}_2$, the boron-bridged path is the shorter one and is labelled $J_{6a}$, whereas in $\mathrm{TaCoB}_2$ it is the longer path and is labelled $J_{6b}$. The apparent reversal of the $J_{6a}$--$J_{6b}$ sign sequence in $\mathrm{TaCoB}_2$ therefore reflects the distance-based labelling convention rather than a change in the underlying exchange mechanism.
\\ The cutoff and exchange-pathway analyses therefore give a consistent microscopic picture. The sixth nearest neighbour introduces two crystallographically distinct inter-sublattice pathways, $J_{6a}$ and $J_{6b}$. Their inequivalence lifts the magnon degeneracy. The relevant exchange scale is therefore the difference between these two couplings, $\Delta J_{\mathrm{AM}}=|J_{6b}-J_{6a}|$, rather than either coupling alone. More distant interactions then provide quantitative corrections to the splitting magnitude. The microscopic origin of the chiral magnon splitting in XCoB$_2$ is therefore the inequivalence of these long-range sixth-neighbor inter-sublattice exchange pathways. 
\subsection{Monte Carlo Determination of the Neel Temperature}To quantify the thermal stability of the compensated (altermagnetic) order in TaCoB$_2$, ZrCoB$_2$, and HfCoB$_2$, we performed classical Monte Carlo (MC) simulations using the calculated intersite exchange parameters $J_{ij}$. The N'eel temperature $T_N$ is determined from the common crossing of the Binder-cumulant curves $U_L(T)$ \cite{Binder1981} and confirmed by the susceptibility peak, which becomes sharper with increasing system size. Figure~\ref{fig:fig-6} shows the results for all three compounds.
\\
Figure~\ref{fig:fig-6}(a) shows this behavior for TaCoB$_2$, with a clear peak that sharpens with increasing $L$, while the corresponding curves in Figure~\ref{fig:fig-6}(b) cross within a narrow temperature range, giving $T_N=67\pm1.1$~K. For ZrCoB$_2$, the peaks in Figure~\ref{fig:fig-6}(c) sharpen around $330$~K, and the crossing in Figure~\ref{fig:fig-6}(d) yields $T_N=330\pm5.7$~K. Similarly, HfCoB$_2$ shows pronounced peaks near $307$~K in Figure~\ref{fig:fig-6}(e), with the corresponding crossing in Figure~\ref{fig:fig-6}(f) giving $T_N=307\pm5.0$~K. The agreement between the two finite-size methods provide consistent estimates of $T_N$, with ZrCoB$_2$ exhibiting the highest ordering temperature, closely followed by HfCoB$_2$, while TaCoB$_2$ orders at a substantially lower temperature.
These ordering temperatures place ZrCoB$_2$ and HfCoB$_2$ among the relatively few altermagnetic candidates with magnetic transition temperatures approaching or exceeding room temperature, which is an important requirement for practical spintronic applications \cite{Smejkal2022a, Song2025NRM,Chumak2015}. For comparison, the semiconducting altermagnet MnTe orders at $T_N\approx310$~K~\cite{Liu2024MnTe}, while the insulating rutile fluorides MnF$_2$, CoF$_2$, and NiF$_2$ order at $67$, $37$, and $73$~K, respectively~\cite{Yuan2020Fluorides}. The metallic altermagnet CrSb exhibits an exceptionally high $T_N\approx700$~K, making it attractive for room-temperature applications~\cite{Reimers2024CrSb}. The extensively studied candidate RuO$_2$ has also been proposed as a room-temperature altermagnet with giant spin splitting~\cite{Smejkal2023chiral}, although its magnetic ground state remains under debate~\cite{Kessler2024RuO2}. Within this landscape, ZrCoB$_2$ ($330$~K) and HfCoB$_2$ ($307$~K) are comparable to MnTe and lie near or above room temperature, making them promising for room-temperature device operation. In contrast, TaCoB$_2$ ($67$~K), with an ordering temperature comparable to that of the rutile fluoride MnF$_2$, would require operation below room temperature.
\\
Finally, the pronounced hierarchy
$T_N^{\mathrm{Ta}} \ll T_N^{\mathrm{Hf}} \lesssim T_N^{\mathrm{Zr}}$
can be understood qualitatively from the calculated exchange interactions.
The dominant nearest-neighbor Co--Co coupling in TaCoB$_2$ is approximately
$5$~meV, substantially smaller than the approximately $18$~meV obtained for
both ZrCoB$_2$ and HfCoB$_2$. This reduction in the dominant exchange scale is
consistent with the much lower $T_N$ of TaCoB$_2$, although the transition
temperature is determined by the full network of exchange interactions. The
similar $T_N$ values of ZrCoB$_2$ and HfCoB$_2$ are also consistent with the
chemical similarity of Zr and Hf, which are isovalent group-4 elements with
comparable atomic sizes due to the lanthanide contraction. In contrast, Ta
belongs to group~5 and introduces an additional valence electron relative to
Zr and Hf, which may modify the electronic states involved in the Co--Co
exchange. A detailed orbital-resolved analysis would be required to establish
the microscopic origin of this difference.
\subsection{Weyl Semimetal Topology}
\begin{figure}
	\centering
	\includegraphics[width=1.0\linewidth]{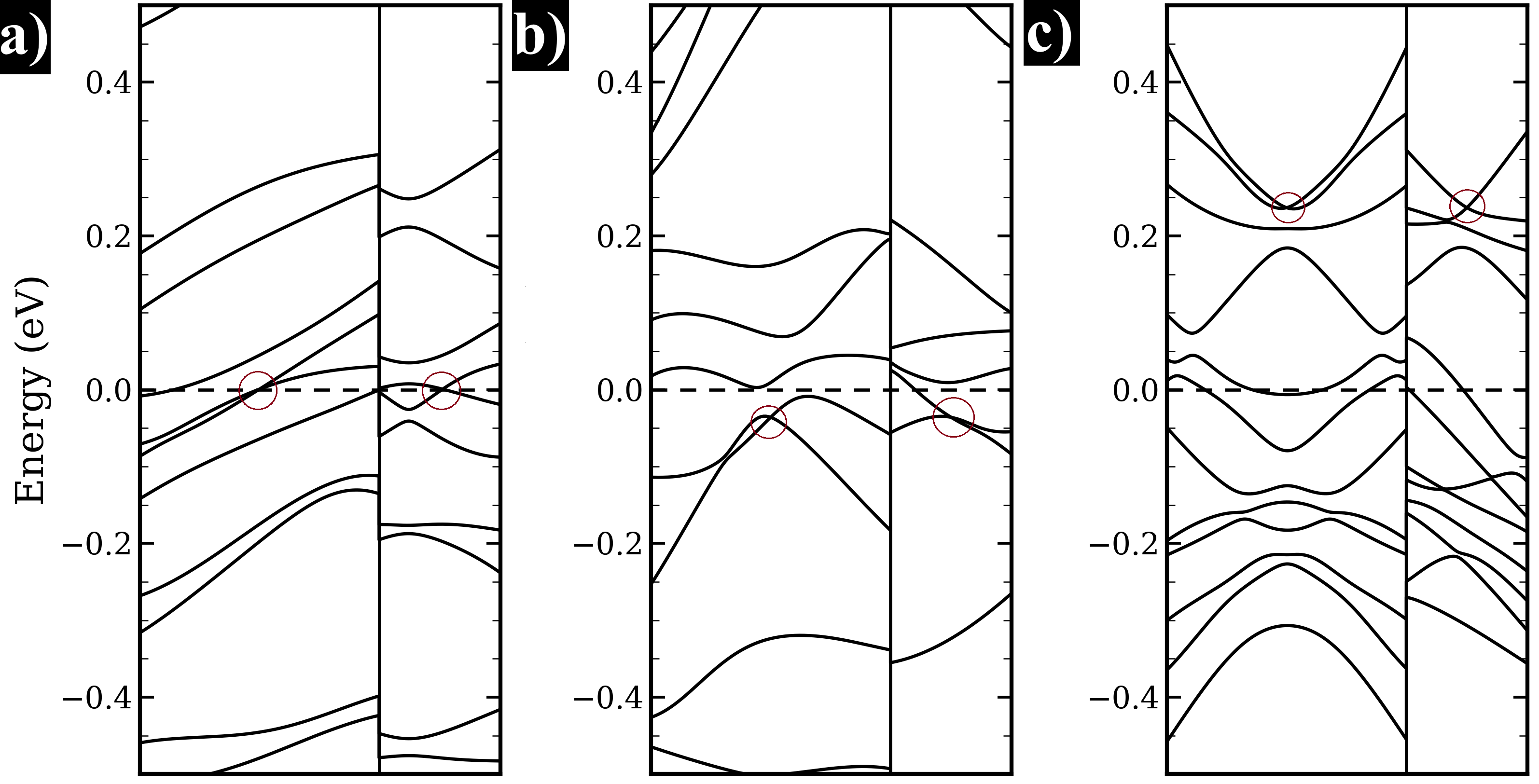}
	\caption{Electronic band structures in the vicinity of the Fermi level for (a) TaCoB$_2$, (b) ZrCoB$_2$, and (c) HfCoB$_2$. The linear band crossings highlighted by the circles correspond to the Weyl nodes. The Weyl nodes are located at the Fermi level in TaCoB$_2$, whereas they occur at $E=-0.036$~eV and $E=0.2376$~eV for ZrCoB$_2$ and HfCoB$_2$, respectively.}
	\label{fig:fig-7}
\end{figure}

Having confirmed the robustness of the altermagnetic phase, we now turn to its topological electronic properties. Figure~\ref{fig:fig-7}(a--c) presents the enlarged electronic band structures in the vicinity of the Fermi level for TaCoB$_2$, ZrCoB$_2$, and HfCoB$_2$, respectively. Distinct linear band crossings, indicative of Weyl nodes, are clearly observed in all three compounds. Notably, the Weyl nodes in TaCoB$_2$ lie exactly at the Fermi level, whereas those in ZrCoB$_2$ and HfCoB$_2$ are located at $E=-0.036$~eV and $E=0.2376$~eV, respectively. To confirm the Weyl semimetallic nature of these compounds, we determined the locations and chiralities of all Weyl nodes using WannierTools. Eight Weyl nodes are identified in each compound, consisting of four pairs of opposite chirality ($\mathscr{C}=\pm1$), consistent with the Nielsen--Ninomiya no-go theorem, which requires the net chiral charge in the Brillouin zone to vanish \cite{nielsen1981absence,nielsen1981absence2,armitage2018weyl}. The Weyl nodes are related by crystal symmetry and are distributed symmetrically throughout the Brillouin zone. The complete coordinates and chiralities of all Weyl nodes for TaCoB$_2$, ZrCoB$_2$, and HfCoB$_2$ are provided in Table~S1 of the Supplementary Information.
\begin{figure*}
	\centering
	\includegraphics[width=1\linewidth]{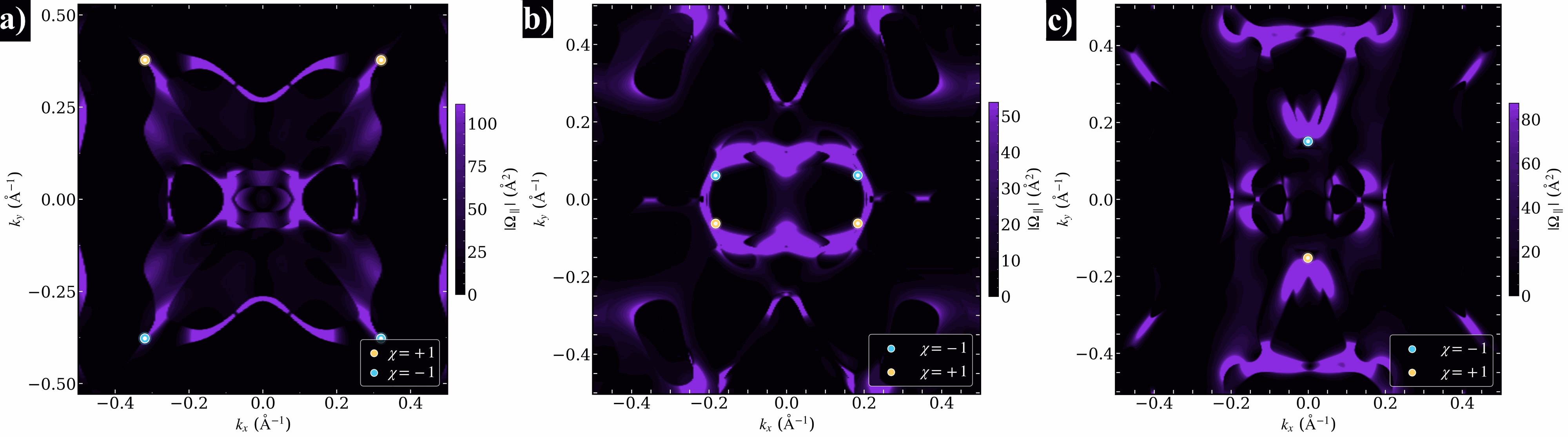}
	\caption{Berry curvature distributions in the $k_x$--$k_y$ plane for (a) TaCoB$_2$, (b) ZrCoB$_2$, and (c) HfCoB$_2$. The projected Weyl nodes are indicated by their chiralities ($\mathscr{C}=\pm1$). The Berry-curvature maxima coincide with the projected Weyl-node positions, reflecting the monopole character of the Weyl nodes and their associated topological charge.}
	\label{fig:fig-8}
\end{figure*}
\begin{figure*}
	\centering
	\includegraphics[width=1.0\linewidth]{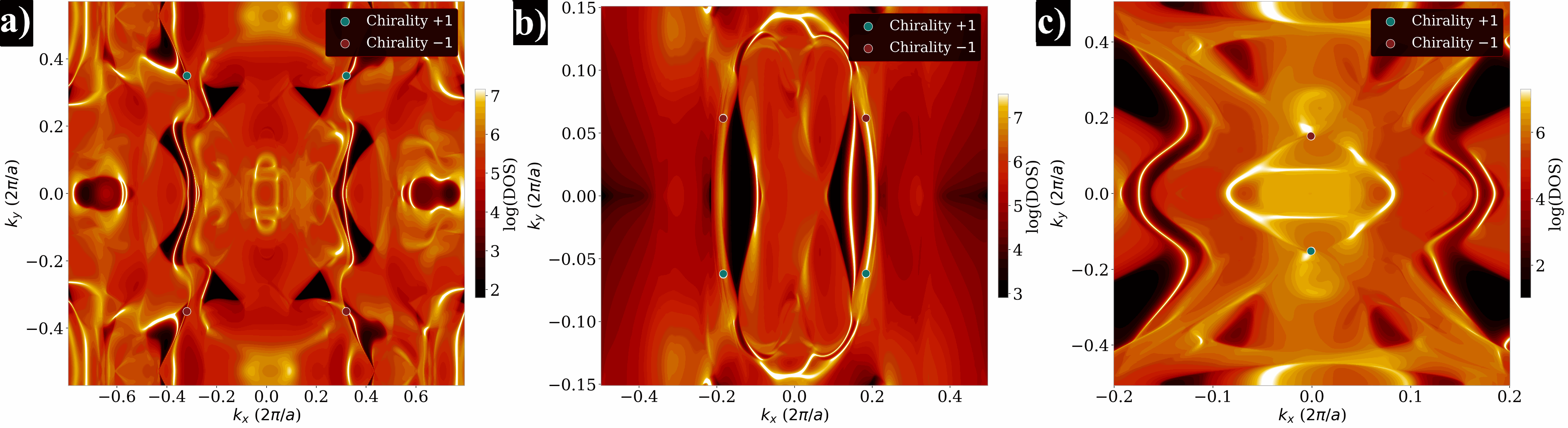}
	\caption{Surface spectral functions of the (001) surface for (a) TaCoB$_2$, (b) ZrCoB$_2$, and (c) HfCoB$_2$ evaluated at the corresponding Weyl-node energies. The projected Weyl nodes are indicated by their chiralities. The termination of the open Fermi arcs at Weyl nodes of opposite chirality reflects the nontrivial bulk topology.}
	\label{fig:fig-9}
\end{figure*}
\\To gain further insight into the topology associated with these Weyl nodes, we computed the Berry curvature distribution in the $k_x$--$k_y$ plane. In Weyl semimetals, the Berry curvature acts as an effective magnetic field in momentum space~\cite{wu1975concept,berry1984proc,bliokh2005spin,cooper2012designing}, with Weyl nodes acting as monopole sources ($Q_{\mathrm{W}}=+1$) and sinks ($Q_{\mathrm{W}}=-1$) of the Berry flux~\cite{wan2011topological}. Figure~\ref{fig:fig-8}(a) shows the Berry curvature distribution of TaCoB$_2$, exhibiting a distinct butterfly-like pattern with mirror symmetry about both the $k_x$ and $k_y$ directions. Pronounced Berry-curvature hot spots are localized at the projected Weyl nodes. Figure~\ref{fig:fig-8}(b) presents the Berry curvature distribution of ZrCoB$_2$, characterized by a fourfold symmetric ring-like feature elongated along the $k_x$ direction around the Brillouin-zone center together with localized hot spots at the projected Weyl nodes. Likewise, Figure~\ref{fig:fig-8}(c) displays the Berry curvature distribution of HfCoB$_2$, exhibiting an hourglass-like feature extending along the $k_y$ direction with the projected Weyl nodes located approximately on the $k_y$ axis ($k_x\approx0$). In all three compounds, pronounced Berry-curvature maxima occur at the projected Weyl-node positions.\\
To quantitatively verify the topological charge of these Weyl nodes, we evaluated the Chern number on a small closed surface enclosing each node. Every Weyl point carries a quantized Chern number of $|C|=1$, confirming elementary Weyl nodes with unit topological charge. The observed Berry curvature and quantized Chern numbers establish the nontrivial bulk topology of all three compounds. The finite Berry curvature can be understood as a consequence of the altermagnetic order. Unlike $\mathcal{PT}$-symmetric antiferromagnets, where the Berry curvature vanishes identically \cite{xiao2010berry,vsmejkal2020crystal}, broken $\mathcal{PT}$ symmetry in altermagnets allows a nonzero Berry curvature, while the remaining crystal and magnetic symmetries constrain its distribution throughout the Brillouin zone.
\\We next investigate the topological surface states by calculating the semi-infinite (001) surface spectral functions using the iterative Green's function method at the corresponding Weyl-node energies. For TaCoB$_2$, Figure~\ref{fig:fig-9}(a) shows distinct open Fermi arcs connecting projected Weyl nodes of opposite chiralities, providing a clear signature of the nontrivial bulk topology. Similar features are observed for ZrCoB$_2$ [Figure~\ref{fig:fig-9}(b)], where the Fermi arcs remain clearly distinguishable despite the significant projected bulk spectral weight around the Brillouin-zone center. Likewise, HfCoB$_2$ [Figure~\ref{fig:fig-9}(c)] exhibits open Fermi arcs connecting the projected Weyl nodes. Although the different projection of the bulk Weyl nodes leads to different Fermi-arc patterns among the three compounds, the characteristic open Fermi arcs remain robust. \begin{figure*}
	\centering
	\includegraphics[width=1.0\linewidth]{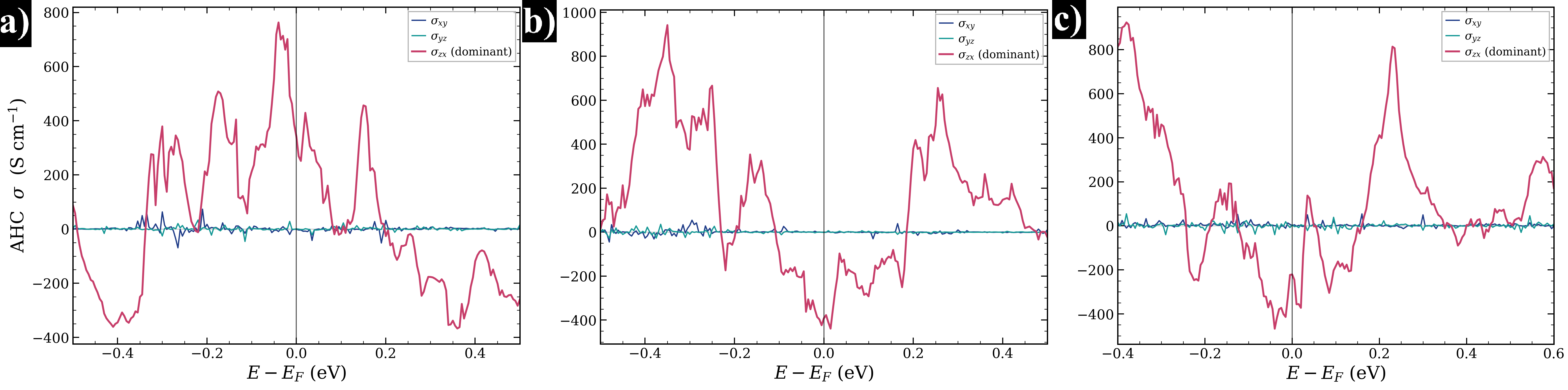}
	\caption{Calculated energy-dependent intrinsic anomalous Hall conductivity tensor components of (a) TaCoB$_2$, (b) ZrCoB$_2$, and (c) HfCoB$_2$. The dominant $\sigma_{zx}$ component is shown together with the much smaller $\sigma_{xy}$ and $\sigma_{yz}$ components. The vertical dashed line indicates the Fermi level ($E-E_F=0$).}
	\label{fig:fig-10}
\end{figure*}
 These topological surface states reflect bulk-boundary correspondence and, together with the Berry-curvature monopoles and quantized Chern numbers, confirm the  Weyl semimetal phase in altermagnetic TaCoB$_2$, ZrCoB$_2$, and HfCoB$_2$.\\
Finally, we examine the intrinsic anomalous Hall conductivity (AHC), a direct transport signature of the Berry curvature generated by the Weyl nodes. As an axial (pseudovector) response, the allowed AHC tensor components are determined by the magnetic point-group symmetry. For the altermagnetic order realized here, only $\sigma_{zx}$ is symmetry-allowed, while $\sigma_{xy}$ and $\sigma_{yz}$ vanish upon integration of the Berry curvature over the Brillouin zone. For TaCoB$_2$, energy-dependent AHC is shown in Figure~\ref{fig:fig-10}(a). The
Hall response is highly anisotropic, with the $\sigma_{zx}$ 
component dominating throughout the investigated energy
range, while $\sigma_{xy}$ and $\sigma_{yz}$ remain close to zero. Since the Weyl nodes lie near the Fermi level, $\sigma_{zx}$ reaches $342$ ~S/cm at $E_F$ and a maximum magnitude of $763$~S/cm at $E-E_F=-0.0398$~eV. Figure~\ref{fig:fig-10}(b) shows a similar response for ZrCoB$_2$, with $\sigma_{zx}$ as the dominant component. Its value is $-392$~S/cm at the Fermi level and $-306$~S/cm at the Weyl-node energy, reaching a maximum magnitude of $942$~S/cm at $E-E_F=-0.350$~eV. Likewise, Figure~\ref{fig:fig-10}(c) shows that the AHC of HfCoB$_2$ is dominated by $\sigma_{zx}$ throughout the investigated energy range, with $-221$~S/cm at the Fermi level and a peak magnitude of $924$~S/cm at $E-E_F=-0.380$~eV. \\The Fermi-level $\sigma_{zx}$ values are comparable to the intrinsic AHC of the compensated Weyl magnet Ti$_2$MnAl ($\sim300$~S/cm)~\cite{Shi2018Ti2MnAl} and exceed those reported for the d-wave altermagnet Mn$_5$Si$_3$ ($5$--$20$~S/cm)~\cite{Reichlova2024Mn5Si3} and the noncollinear antiferromagnet Mn$_3$Sn ($\sim100$~S/cm)~\cite{Nakatsuji2015Mn3Sn}. The peak values of $760$ to $940$~S/cm approach that of the ferromagnetic Weyl semimetal Co$_3$Sn$_2$S$_2$ ($1130$~S/cm)~\cite{Liu2018Co3Sn2S2}. Their proximity to the Fermi level suggests that these large responses may be experimentally accessible through chemical doping or electrostatic gating, as demonstrated in related magnetic systems~\cite{He2024ClCo3Sn2S2,Shen2022SbCo3Sn2S2,Deng2018Fe3GeTe2}. Combined with the vanishing net magnetization of altermagnets, these results make TaCoB$_2$, ZrCoB$_2$, and HfCoB$_2$ promising platforms for low-power, high-density spintronics~\cite{Baltz2018AFM,Jungwirth2016AFM,Smejkal2018TopoAFM,Smejkal2022b}.
\section{Conclusions}
We have investigated the orthorhombic ternary borides $X$CoB$_2$ ($X=$ Ta, Zr, Hf) using symmetry analysis and first-principles calculations, and have established them as a family of metallic $d_{yz}$-wave altermagnets. Symmetry analysis within the magnetic space group $Pnm'a'$ (BNS No.~62.447) predicts a nonrelativistic spin splitting proportional to $k_yk_z$, with degeneracy enforced on the $k_y$ and $k_z$ planes. The calculated band structures reproduce this pattern, with strong spin splitting in the interior of the $k_x=0$ face. The momentum-averaged splitting on this plane reaches 88.4, 52.3, and 70.6~meV at the Fermi level and exceeds 200~meV at its maximum in all three compounds.
\\Interactions up to the fifth nearest neighbour set the magnetic background and the spin-wave energy scale but do not produce an appreciable splitting. The splitting emerges only when the sixth-neighbour inter-sublattice exchange is
	included, which symmetry separates into two crystallographically inequivalent,
	nearly equidistant Co--Co paths. These have opposite signs, namely a ferromagnetic
	Co--B--B--Co channel and an antiferromagnetic Co--B--$X$--Co channel. The
	corresponding exchange inequivalence, $\Delta J_{\mathrm{AM}}=1.01$, $1.80$, and
	$1.92$~meV, increases from TaCoB$_2$ to ZrCoB$_2$ to HfCoB$_2$. The maximum
	chiral magnon splitting follows the same trend, increasing from $1.51$ to
	$2.03$ and $3.10$~meV, respectively. This parallel trend suggests that
	$\Delta J_{\mathrm{AM}}$ may govern the magnitude of the chiral magnon splitting. These splitting values are comparable to those reported for $\alpha$-MnTe and hematite, while occurring below 50~meV. Monte Carlo simulations give N\'eel temperatures of $67\pm1.1$, $330\pm5.7$, and $307\pm5.0$~K, placing ZrCoB$_2$ and HfCoB$_2$ above room temperature.
\\
With spin-orbit coupling included, all three compounds host symmetry-protected
Weyl points of unit Chern number close to the Fermi level, together with the
associated Fermi-arc surface states. These topological features give rise to a
	strong intrinsic anomalous Hall response. XCoB$_2$ ($X=\mathrm{Ta}$, Zr, and Hf)
	therefore provides a compensated platform carrying both magnonic and electronic
	chirality, one in the spin waves and the other in the Berry curvature, with no
	net moment and no stray field.\\These results establish XCoB$_2$ as a rare lower-symmetry three-dimensional metallic altermagnet. The coexistence of chiral magnon splitting and symmetry-protected Weyl states, together with above-room-temperature magnetic ordering in ZrCoB$_2$ and HfCoB$_2$, highlights the distinct properties of this material family. The magnon splitting varies with the $X$ site and follows the same trend as exchange inequivalence $\Delta J_{\mathrm{AM}}$, suggesting that the intermediate bonding network may influence its magnitude. The occurrence of these split magnon branches at low energies also makes them accessible to inelastic neutron scattering and relevant for field-free magnon transport. Given the established synthesis of ternary borides with the XYB$_2$ structure type, we hope that these results will encourage experimental investigation of the predicted chiral magnon and transport signatures.
\section{Acknowledgement} The authors gratefully acknowledge the Department of Physics, University of Dhaka, for its academic and research support. We are also grateful to the Bangladesh Research and Education Network (BdREN) for providing server resources and computational infrastructure that facilitated the simulations and data analysis. This work was further supported by the UNESCO--TWAS Research Grant Programme (Grant No.~24-362 RG/PHYS/ASJ--FR 3240339192), funded by the Swedish International Development Cooperation Agency (Sida), which provided access to high-performance computing (HPC) facilities essential for this work.
\section{Data Access Statement} The data generated and analysed during the current study are available from the corresponding author upon reasonable request. \section{Competing Interests} The authors declare no competing interests.
\bibliography{ref}    
\bibliographystyle{apsrev4-2}
\end{document}